\documentclass[superscriptaddress,amsmath,amssymb,preprint,11pt]{revtex4}

\usepackage{graphicx}% Include figure files
\usepackage{dcolumn}% Align table columns on decimal point
\usepackage{bm}% bold math
\usepackage{setspace} 
\newcommand{\DP}[2]{\ensuremath{\frac{\partial{#1}}{\partial{#2}}}}

\newcommand{\DT}[2]{\ensuremath{\frac{\mathrm{d}{#1}}{\mathrm{d}{#2}}}}

\newcommand{\Tr}[1]{\mbox{Tr$\,{#1}$}}

\usepackage{color}

\begin{document}

\title{From cells to tissue:~A continuum model of epithelial mechanics}

\author{Shuji Ishihara}
\email{csishihara@g.ecc.u-tokyo.ac.jp}
\affiliation{Graduate School of Arts and Sciences, The University of Tokyo, Tokyo 153-8902, Japan}
\affiliation{Department of Physics, School of Science and Technology, Meiji University, Kanagawa 214-8571, Japan}

\author{Philippe Marcq}
\affiliation{Sorbonne Universit\'es, UPMC Universit\'e Paris 6, 
Institut Curie, CNRS, UMR 168, Laboratoire Physico Chimie Curie, 
Paris, France}

\author{Kaoru Sugimura}
\affiliation{Institute for Integrated Cell-Material Sciences (WPI-iCeMS), Kyoto University, Kyoto 606-8501, Japan}
\affiliation{JST PRESTO, Tokyo 102-0075, Japan}%

%\date{\today}

%\setstretch{1.2}

\begin{abstract}
A continuum model of epithelial tissue mechanics was formulated
using cellular-level mechanical ingredients and cell morphogenetic processes,
including cellular shape changes and cellular rearrangements.
This model can include finite deformation, and
incorporates stress and deformation tensors,
which can be compared with experimental data. 
Using this model, we elucidated dynamical behavior
underlying passive relaxation, active contraction-elongation, and tissue shear flow. 
This study  provides an integrated scheme for the understanding 
of the mechanisms that are involved in orchestrating the morphogenetic processes in individual cells,
in order to achieve epithelial tissue morphogenesis.
%% Please note that, generally, the final paragraph of the abstract section should include the 
%% explanation about the importance of your findings and the implications for the field in general. 
%% Please consider including this before the final submission of the manuscript. 
\end{abstract}
\maketitle

\section{Introduction}

\begin{figure}[p]
\centering
 \resizebox{0.47\textwidth}{!}{%
  \includegraphics{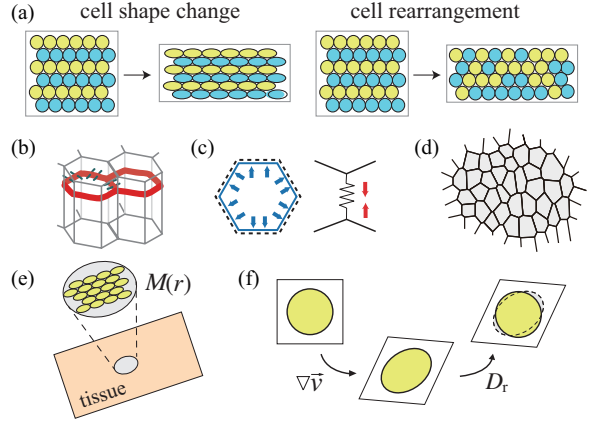}
  }
  \caption{\label{fig:FigModel}
  \textbf{Schematic illustration of the model of epithelial mechanics.}
  (a) Tissue deformation based on cellular shape deformation (left) and by cellular rearrangement (right).
  (b) In epithelial tissues, 
cells adhere to each other via E-Cadherin (green) at adherens junctions (AJ), and acto-myosin (red) runs along the cell junctions.
  (c) Cell pressures (left) and cell junction tensions (right) act in the AJ plane and
   determine epithelial cell shapes.
  (d) CVM schematic representation. Each cell is represented by a polygonal contour.
  (e, f) Schematic illustration of the model presented here.
  (e) A coarse-grained cellular shape tensor $M$ represents the tissue-scale cellular shape field in our model. 
  (f) Kinematics of cellular shape deformation. Cellular shape alterations through tissue deformation $\nabla \vec{v}$ and cell rearrangement $D_{\mathrm{r}}$.
  }
\end{figure}

During tissue morphogenesis, 
tissues acquire their unique shape and size through a series of deformations.
Morphogenesis occurs at multiple levels, and molecular, cellular, and tissue level changes are interdependent.
At cellular level, tissue deformation is accounted for by changes in cell shape, position, and number (Fig.~\ref{fig:FigModel}(a);
hereafter named cell morphogenetic processes), which are triggered by biochemical signaling and 
forces generated by cells \cite{Heisenberg2013, Sampathkumar2014, Guillot2013}.
While the tissue stress can affect cell morphogenetic processes through the changes in molecular activity and localization,
cell morphogenetic processes generate stress \cite{Blanchard2011, Aigouy2010, Sugimura2013, Aw2016, Uyttewaal2012, Schluck2013, Wyatt2015}.
However, the mechanisms by which the shape of a tissue emerges from these multi-scale feedback processes remain unclear.

In order to clarify this, a coarse-grained description and modeling of cellular and tissue dynamics at an appropriate length scale is required:
while the position and timing of cell morphogenetic processes are stochastic at single cell level,
the averaging of values obtained over a larger length scale yields a smooth spatial pattern
that is reproducible among different samples. 
We previously determined the appropriate averaging length scale for 
describing
epithelial tissue dynamics (several tens of cells in a patch), 
and developed coarse-grained methods for measuring stress and kinematics
\cite{Bosveld2012, Guirao2015, Ishihara2012, Ishihara2013}. 
A force inference method was used for the quantification of cell junction tensions and cell pressures (Fig.~\ref{fig:FigModel}(b), (c)),
which can be integrated to calculate a stress tensor \cite{Ishihara2012, Ishihara2013, Chiou2012, Brodland2014}.
A texture tensor method was used for the measuring of different cell morphogenetic processes
(\emph{e.g.}, cell division, cell shape changes) in the same physical dimension,
which can be further integrated to obtain tissue scale, spatio-temporal maps of
tissue growth and cell morphogenetic processes \cite{Guirao2015}. 
Together, these methods provide the information on the amplitude, orientation, and anisotropy of tissue stress,
tissue growth, and cell morphogenetic processes, and correlations between them \cite{Guirao2015}.

A modeling scheme capable of accommodating the quantitative data described above is still lacking \cite{Khalilgharibi2016}. 
Cell-based models, such as the cell vertex model (CVM) \cite{Honda1983} and the cellular Potts model (CPM) \cite{Graner1993} 
are often employed for the simulation of epithelial tissue morphogenesis 
(Fig.~\ref{fig:FigModel}(d); \cite{Fletcher2014,Maree2007,Brodland2004}), and have proven
useful for including experimental data obtained at cellular level, such as the laser ablation of cell junctions or subcellular distribution of proteins. 
However, the relationship between cell morphogenetic processes and tissue scale deformation and 
rheology emerges from numerical simulations without being directly tractable.
Continuum models allow the in-depth analysis of tissue rheology \cite{Fung1993, Tlili2015},
yet in many cases do not include the information of the cellular structure by construction, and thus 
fail to discriminate between different cell morphogenetic processes.
A limited number of studies considered the degrees of freedom that represent cell morphogenetic processes and cell polarity
\cite{Shraiman2005, Ranft2010, Tlili2015, Popovic2016}, 
but do so in the context of macroscopic models, which do not incorporate cell-level mechanical parameters explicitly.
The finite-element model introduced in \cite{Brodland2006} includes at
a coarse-grained level the contributions of cellular rheology, shape changes,
rearrangements and divisions.
A continuum model has been derived from the CVM previously 
but without considering cell rearrangements \cite{Murisic2015} (see also \cite{Turner2005, Fozard2010} in 1D). 

The main aim of our study was to develop a two-dimensional hydrodynamic model of the epithelial tissue.
This model included a field that represents coarse-grained cell shape, which 
enabled us to treat different types of cell morphogenetic processes 
distinctively.
First, kinematics was identified by decomposing tissue deformation into cell shape changes and cell rearrangements.
Following this, by introducing an energy function deduced from CVM/CPM, thermodynamic formalism was employed to determine kinetics.
The model we derived here describes tissue deformation through stress and deformation tensors,
which can be compared with the data obtained experimentally \cite{Guirao2015, Ishihara2012, Ishihara2013}, and can incorporate active terms smoothly.
We solved the model for several conditions typical for deforming planar tissues during development,
and demonstrated that the model predicts the relaxation of cellular shape following the tissue stretching,
the relation of the direction of cell elongation and tissue flow during active contraction-elongation (CE), and shear-thinning in tissues.
Our approach provides a theoretical framework that enables to assess how cellular level 
mechanical parameters and cell morphogenetic processes are integrated to realize tissue-scale deformation.

\section{Model}

\subsection{Cellular shape tensor}
To construct a continuum model, we approximated each cell by an ellipse, characterized by a $2 \times 2$ symmetric tensor $M$.
Cellular shape can be expressed as
$(\vec{r}-\vec{r}_c)^T M^{-1}(\vec{r}-\vec{r}_c) = 1$,
where $\vec{r}_c$ represents the center of a cell and superscript~$^T$ denotes the transpose
(Fig.~\ref{fig:FigModel}(e)).
Since the eigenvalues of $M$ are the square lengths of ellipse semi-axes, cell area can be presented as $A = \pi |M|^{1/2}$, where $|M|$ is the determinant of $M$.
By coarse-graining over a representative surface element
comprising a sufficient number of cells \cite{Bosveld2012, Guirao2015},
we obtained a spatially smooth tensor field $M(\vec{r})$.
Similar to a previously described texture tensor \cite{Guirao2015},
the symmetric tensor $M(\vec{r})$ represents measure of tissue scale deformation
in our model, with the physical dimension of square length,
which can be experimentally quantified from %the 
segmented images of two-dimensional (2D) epithelia.

%%%%% Kinematics %%%%%%%
\subsection{Kinematics}
Total tissue deformation rate can be represented by the tensor $\nabla \vec{v}$, in which $\vec{v}$ is the velocity field. Here, we used $\left( \nabla \vec{v} \right)_{ij} = \partial_j v_i$,
where indices $i$ and $j$ denote cartesian coordinates.
The deformation rate $\nabla \vec{v}$ represented the sum of its symmetric part $D = (\nabla \vec{v}+[\nabla \vec{v}]^T)/2$ and its antisymmetric part
$\Omega = \left( \nabla \vec{v}- [\nabla \vec{v}]^T\right)/2$.
We decomposed tissue deformation rate into the sum 
of contributions, due to the cellular shape alterations and other cell morphogenetic processes, 
and here, we considered cell rearrangement, division, and death:
\begin{equation}
	\nabla \vec{v} = \Omega + D_{\mathrm{s}} + D_{\mathrm{r}} \,. 
	\label{eq:deformation_dec}
\end{equation}
The quantity $\Omega + D_{\mathrm{s}}$ 
represents the tissue deformation rate stemming from cellular shape alterations,
while $D_{\mathrm{r}}$ denotes the deformation rates
that involve topological changes in a network of cell junctions, 
\emph{i.e.} cell rearrangement, division, and death. 
We assumed that these processes are irrotational, 
so that $D_{\mathrm{r}}$ is symmetric.
In practice, these tensors can be experimentally determined by cellular shape tracking \cite{Guirao2015}.	
In cellular materials, $D_{\mathrm{r}}$ is kinematically associated with cell shape changes \cite{Tlili2015}:
\begin{equation}
\dot{M} - (\nabla \vec{v} - D_r ) M - M (\nabla \vec{v} - D_r )^T = 0 \,,
\label{eq:Kinematics}
\end{equation}
where $\dot{M} \equiv \partial_t M + \vec{v} \nabla M$
represents the Lagrange derivative of $M$ (Fig.~\ref{fig:FigModel}(f)).
The kinematic relationship \eqref{eq:Kinematics}
was derived for the non-affine deformation in the rheology of polymer melts \cite{Leonov1976,Leonov1987,Larson1988} and
foams \cite{Marmottant2008, Benito2008},
where $D_{\mathrm{r}}$ can be interpreted as due to the slippage 
between polymer molecules and between bubbles, respectively. 
When $D_{\mathrm{r}} = 0$, the left-hand side of \eqref{eq:Kinematics} becomes 
the co-deformational upper-convected derivative \cite{Larson1983}.
In SI Section~3, we demonstrate that the conservation of cell number density 
$\rho = 1/\pi |M|^{1/2}$ can be presented as 
$\partial_t \rho + \nabla \cdot (\rho \vec{v}) = (\Tr{D_{\mathrm{r}}})\rho$,
where $\Tr{}$ denotes the trace, and
$\Tr{D_{\mathrm{r}}}$ coincides with the variation rate of $\rho$.
The effects of cell division and cell death should be investigated in future studies, while here we considered a situation in which individual cells only deform
elastically and/or intercalate, and the tissue area is invariant and the deformation
rate is traceless ($\Tr{D_{\mathrm{r}}} = 0$).%; Fig.~\ref{fig:FigModel}(a)).%).

\subsection{Energy function and elastic stress}
In CVM and CPM, cell geometry can be determined by minimizing energy function \cite{Fletcher2014, Farhadifar2010} 
\begin{equation}
f_{c} =  \sum_i \frac{K}{2}\left(A_i-A_0\right)^2 + \sum_{[ij]} \gamma_0 \ell_{ij} + \sum_i\frac{\kappa_0}{2}L_i^2 \,,
\label{eq:cell_energy}
\end{equation} 
where $A_i$ and $L_i$ represent the area and perimeter, respectively,
of cell $i$, and $\ell_{ij}$ is the length of the interface between cells $i$ and $j$.
The first term represents cell-area elasticity with elastic modulus $K$
and reference cell area $A_0$.
The second and third terms represent cell junction tensions,
where the tension is given by $\gamma_0 + \kappa_0 (L_i + L_j)$.

Here, by using the cell shape tensor $M(\vec{r})$,
we considered an energy function for the continuum model, which is comparable 
with that of the CVM/CPM, \eqref{eq:cell_energy}.
For arbitrary semi-axes $a$, $b$, the perimeter of an ellipse can be given
by Euler's formula (see Suppl. Fig. S1) \cite{Chandrupatla2010}:
\begin{equation}
  \label{eq:Euler}
  L(a,b) = \pi \, \sqrt{2 (a^2 + b^2)} \, 
\mathcal{F}\left( \frac{1}{4}, -\frac{1}{4}; 1; h^2 \right), 
\, h = \frac{a^2-b^2}{a^2+b^2} \,,
\end{equation}
where $\mathcal{F}$ is a hypergeometric function. 
Using the Cayley-Hamilton equation, we derived the identities $a^2 + b^2 = \Tr{M}$,
$(a^2 - b^2)^2 =  \Tr{M}^2 - 4 |M|$.
Upon coarse-graining \eqref{eq:cell_energy}, we obtained energy 
density per unit area
\begin{equation}
\label{eq:F:anisotropic}
F = \frac{1}{\pi|M|^{1/2}} 
 \left[  \frac{K}{2} \left( \pi |M|^{1/2} - A_0\right)^2 
+ \frac{\gamma_0}{2} L(M) +\frac{\kappa_0}{2}\!\ L(M)^2
\right]   \,, 
\end{equation} 
where $L(M)$ is defined for arbitrarily large cellular shape anisotropy by
\begin{equation}
  \label{eq:Euler:M}
  L(M) = \pi \, \sqrt{2 \Tr{M}} \, 
\mathcal{F}\left( \frac{1}{4}, -\frac{1}{4}; 1; 
1 - \frac{4 |M|}{ (\Tr{M})^2}\right) \,.
\end{equation}
The total elastic energy was obtained by integration
$\mathcal{F} = \int F(M(\vec{r})) \, \mathrm{d}\vec{r}$.

We focused, for simplicity, on conditions close
to the isotropic case (see SI Sec.~1.2 for higher-order approximations).
Expanding $\mathcal{F}$ close
to $a = b$, or $(\Tr{M})^2 = 4 |M|$, the zero order energy can be written as
\begin{equation}
  \label{eq:F:almost:isotropic}
F \!=\! \frac{1}{\pi|M|^{1/2}} \!\!
\left[ \frac{K}{2} \left( \pi |M|^{1/2} - A_0\right)^2 
+ \frac{\pi \gamma_0}{\sqrt{2}} \sqrt{\Tr{M}} 
+ \pi^2 \kappa_0 \Tr{M}
\right] \,,
\end{equation} 
from which the elastic stress can be further derived as
\begin{equation}
\label{eq:Sigma_e}
\sigma_{\mathrm{e}} =    K(\pi |M|^{1/2} -A_0)I + 
\left(
\frac{\gamma_0}{\sqrt{2 \Tr{M}}}
+  2 \pi \kappa_0 
\right) \, \frac{M}{|M|^{1/2}}  \,,
\end{equation}
where $I$ is the unit tensor 
(SI Sec.~1, general case).
The first and second terms represent the
isotropic pressure $-P^{\mathrm{ce}}I$ 
due to the area-elasticity of cells,
and the cellular shape-dependent stress $\sigma^{\mathrm{T}}$ 
due to cell junction tensions, respectively. 
\eqref{eq:Sigma_e} is consistent with the expression of the
Batchelor stress tensor \cite{Batchelor1970, Ishihara2012, Ishihara2013} 
relating tissue-scale stress to cell pressures and cell junction tensions.
Note that $\sigma_{\mathrm{e}}$ and $M$ commute, and therefore, share the same eigendirections, which is consistent with our previous observation
showing that cells are elongated along the inferred maximal stress direction 
in \textit{Drosophila} epithelia \cite{Sugimura2013, Guirao2015, Ishihara2012}.

\begin{figure}[p]
\centering
 \resizebox{0.49\textwidth}{!}{%
   \includegraphics{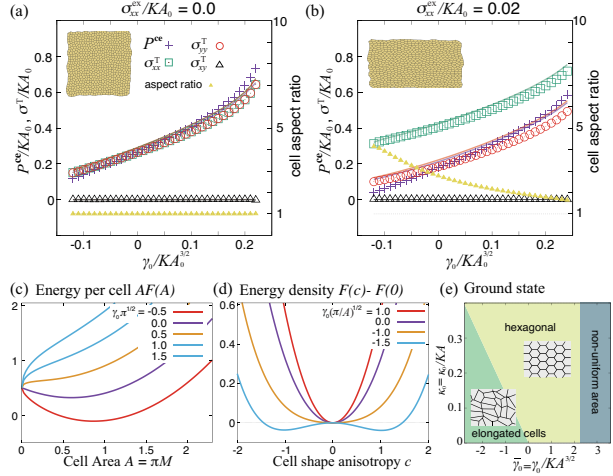}
  }
  \caption{ \label{fig:Potential}
\textbf{Energy function and elastic stress.} 
 (a,b) Macroscopic stress expressions calculated from coarse-grained cellular 
shape tensor $M$ (symbols) are compared with the true ones (solid lines) 
obtained using the CVM simulations (left vertical axis), as a function of
the non-dimensional parameter $ \gamma_0/KA_0^{3/2}$, with $\kappa_0/KA_0 = 0.04$. 
$P^{\mathrm{ce}}$  and  $\sigma^{\mathrm{T}}$ represent the pressure
stemming from cell elasticity and the stress generated by cell junction tensions, respectively. Yellow triangles denote the mean cellular shape aspect ratio (right vertical axis), equal to $e^{2c}$ in terms of the cell shape anisotropy. 
The external stress was set as $\sigma^{\mathrm{ex}}_{xy} =\sigma^{\mathrm{ex}}_{yy} = 0$, with (a) $\sigma^{\mathrm{ex}}_{xx} = 0$ (isotropic case) and (b) $\sigma^{\mathrm{ex}}_{xx} \ne 0$ (anisotropic case).
  (c) Energy per cell as a function of the cell area $A$ at 
	$\gamma_0 \pi^{1/2} = -0.5, 0.0, 0.5, 1.0$ 
	and $1.5$ with  $K = 1.0$  and $2 \pi \kappa_0 = 0.4$.
  (d) Energy density function $F(c)$ at $\gamma_0(\pi/A)^{1/2} = 1.0, 0.0, -1.0$, and $-1.5$ with $2  \pi \kappa_0/A = 0.6$.
  (e) A ground state phase diagram is shown as a function of non-dimensional parameters 
  $\overline{\gamma}_0 = \gamma_0/KA^{3/2}$ and $\overline{\kappa}_0 = \kappa_0/KA$, defined using a cell area 
  $A = \pi M_0$, instead of $A_0$. %\cite{Farhadifar2010,Staple2010}.  
}
\end{figure}

We performed numerical simulation of CVM under the isotropic and 
anisotropic stress environments, and compared the coarse-grained stress 
with the true one. Coarse-grained cell shape $M$ was evaluated by
averaging the second moment of cell shape, and Euler expansion
up to the second order was considered (SI Sec.~1.2). 
The results obtained here confirmed that the coarse-grained stress values agree with those obtained for the true one
(Fig.~\ref{fig:Potential}(a,b), Suppl. Fig. S2). 
 
Factorizing the tensor $M$ as $M = M_0e^{c\Theta}$ can be useful \cite{Merkel2017},
and here, $M_0$ and $c$ are scalar fields
and $\Theta$ is a symmetric, trace-less tensor field 
parameterized by the angle $\theta$ as
 \begin{equation}
 \Theta = 
 \begin{pmatrix}
 \cos 2\theta & ~\sin 2\theta \\
 \sin 2\theta & -\cos 2\theta
 \end{pmatrix}\,.
 \end{equation}
Since $\Theta^2 = I$, we deduced $M = M_0 [\cosh(c) I + \sinh(c) \Theta]$, 
where $M_0$ quantifies the coarse-grained cell area 
$A = \pi |M|^{1/2} = \pi M_0$, dimensionless parameter $c$ 
characterizes the coarse-grained cell shape anisotropy,
and the angle $\theta$ represents the direction of the major axis 
of ellipses. Since $\Tr{M} = 2M_0 \cosh(c)$, $L(M)$ and $F(M)$ 
depend only on $M_0$ and $c$.

For the energy function presented in \eqref{eq:F:almost:isotropic},
the energy per cell, $A\,F(A,c=0)$, is shown 
as a function of $A$ in Fig.~\ref{fig:Potential}(c)
in the isotropic case $c = 0$. For
large values of $\gamma_0$, the functional form becomes concave, 
indicating thermodynamic instability of the state of homogeneous cell area.
Fig.~\ref{fig:Potential}(d) shows $F(A,c)$
as a function of $c$ at constant cell area. 
Circular cell shape ($c=0$) becomes unstable 
for sufficiently large negative values of $\gamma_0$, 
where cells no longer prefer a hexagonal configuration, 
but adopt an elongated shape. 
We recovered two instabilities described for the CVM
\cite{Farhadifar2010, Staple2010} (Fig. \ref{fig:Potential}(e); SI Sec.~1.3),  
showing that the tissue scale energy density $F(M)$ retains the essential 
features of the original cell-based models.

%%%%% Energy function and elastic stress %%%%%%%
%%%%%%%%%%%%%%%

%%%%% Kinetics %%%%%%%
\subsection{Thermodynamic formalism}
Since existing cell-based models, including CVM and CPM,
use \emph{ad hoc} prescriptions for kinetics, we considered the thermodynamic 
formalism \cite{Groot1962} in order to derive generic hydrodynamic equations. 
The total stress tensor $\sigma$ can be
given by the sum of elastic and dissipative stresses, as 
\begin{equation}
  \label{eq:totalstress}
 \sigma = \sigma_{\mathrm{e}} + \sigma_{\mathrm{p}} \,.
\end{equation}
The entropy production rate of an isothermal process was calculated as 
\cite{Leonov1976,Leonov1987}:
\begin{equation}
	T\dot{s} = \sigma\!:\! D - \sigma_{\mathrm{e}}\!:\! D_{\mathrm{s}} = \sigma_{\mathrm{p}} \!:\!D + \sigma_{\mathrm{e}}\!:\!D_{\mathrm{r}}   
\label{eq:EntropyProd} \,,
\end{equation}
where $s$ is the entropy density, and $T$ is the temperature.
Here, %$a\!\!:\!\!b$ 
$a\!\!:\!\!b \equiv \Tr{[ab^T]}$ denotes the scalar product of two arbitrary 
tensors $a$ and $b$,
and $a' \equiv a \!-\! (\Tr{a}/2) I$ is the deviatoric part of $a$.
Because $D_{\mathrm{r}}$ is traceless, we can replace $\sigma_{\mathrm{e}}$ by 
${\sigma_{\mathrm{e}}}'$ in \eqref{eq:EntropyProd}.
By identifying conjugate flux-force pairs as $\sigma_{\mathrm{p}}$-$D$ and $D_{\mathrm{r}}$-${\sigma_{\mathrm{e}}}'$,
the fluxes $(\sigma_{\mathrm{p}}, D_{\mathrm{r}})$ 
can be given by using the linear combinations of the forces $(D,{\sigma_{\mathrm{e}}}')$:
\begin{align}
\sigma_{\mathrm{p}} &= \chi^{\mathrm{ss}} D - \chi^{\mathrm{sr}} {\sigma_{\mathrm{e}}}' \,,\label{eq:Sp}\\
D_{\mathrm{r}} &= \chi^{\mathrm{rs}} D + \chi^{\mathrm{rr}} {\sigma_{\mathrm{e}}}'  \,. \label{eq:Drc}
\end{align} 
The coefficients $\chi^{\mathrm{ss}}, \chi^{\mathrm{sr}}, \chi^{\mathrm{rs}}$, and $\chi^{\mathrm{rr}}$ are fourth-order tensors that satisfy 
Onsager's reciprocity (\emph{e.g.}, $\chi^{\mathrm{sr}}_{ijkl} = \chi^{\mathrm{rs}}_{klij}$).
Note that Maxwell's model was obtained for 
$\chi^{\mathrm{ss}} = \chi^{\mathrm{sr}} = 0$ 
($\sigma = \sigma_{\mathrm{e}}$, $\sigma_{\mathrm{p}} = 0$), 
and that Kelvin-Voigt's model was obtained for 
$\chi^{\mathrm{rs}} = \chi^{\mathrm{rr}} = 0$ ($D = D_{\mathrm{s}} $, 
$D_{\mathrm{r}} = 0$) \cite{Leonov1976,Leonov1987}. 
The term $\chi^{\mathrm{ss}}D$ characterizes dissipative stress due 
to the tissue strain rate, and reduces it to the usual bulk and shear 
viscous terms for isotropic material.
According to \eqref{eq:Drc}, cell rearrangements may be driven both by
the tissue strain rate and by its elastic stress 
\cite{Aigouy2010, Sugimura2013, Etournay2015}. 
The presence of the cross-term  
$- \chi^{\mathrm{sr}} {\sigma_{\mathrm{e}}}'$ in \eqref{eq:Sp} 
is a non-trivial prediction of the model, which cannot be represented 
by a rheology diagram in terms of a combination of springs and dashpots.

In the absence of external forces, the above equations can predict
the relaxation to the steady state, and are not sufficient to address the
active phenomena, such as the sustained epithelial flow \cite{Sato2015} or
self-organized spatio-temporal patterning \cite{Saw2017}.
Therefore, we included
the formalism of active gels \cite{Kruse2005,Marchetti2013},
and added the term $r\Delta \mu$ to the entropy production rate 
\eqref{eq:EntropyProd}:
\begin{equation}
	T\dot{s} = \sigma_{\mathrm{p}} \!:\!D + {\sigma_{\mathrm{e}}}'\!:\!D_{\mathrm{r}}   + r \, \Delta \mu \,,
\label{eq:EntropyProd:active} 
\end{equation}
where $\Delta \mu$ represents the change in chemical 
potential associated with a chemical reaction that supplies energy to the system, and $r$ is the reaction rate. 
By identifying $r$-$\Delta \mu$ as an additional force-flux pair, 
additional terms $\sigma_a$ and $D_a$,
which we refer to as active stress and active cell rearrangement, respectively, 
appeared in \eqref{eq:Sp} and \eqref{eq:Drc}, both of which are proportional to $\Delta \mu$.

The coupling coefficients can depend on $M'$.
Including lowest-order non-linearities, with the condition that $D_{\mathrm{r}}$ is traceless, the generic form of the force-flux relationships can be written as:
\begin{align}
\sigma_{\mathrm{p}} &=   \eta \, D' + \eta' \, (\Tr{D})I \nonumber\\ 
%& + \mu \, (D M'+ M' D -\Tr{(D M')}I)  + \mu' \, (\Tr{D}) M'
 &  + \mu \, (D M'+ M' D)  + \mu' \, (\Tr{D}) M' + \mu'' (\Tr{(D M')})I 
\nonumber \\
& -\nu_1 {\sigma_{\mathrm{e}}}' -\nu_2 \left( {\sigma_{\mathrm{e}}}' M' + M' {\sigma_{\mathrm{e}}}'  \right) 
-\nu_3 ({\sigma_{\mathrm{e}}}'\!:\!M')I \nonumber \\
& - \zeta_1 \Delta \mu \, I - \zeta_2 \Delta \mu \, M'
\,, \label{eq:SigmaP}\\
D_{\mathrm{r}} &= \nu_1 D' + \nu_2 (D M'+ M' D-\Tr{(D M')}I) + \nu_3 (\Tr{D}) M' \nonumber \\
&+ \eta^{-1}_1 {\sigma_{\mathrm{e}}}' +\eta^{-1}_2 \left( {\sigma_{\mathrm{e}}}' M' + M' {\sigma_{\mathrm{e}}}'- \Tr{({\sigma_{\mathrm{e}}}' M')}I \right) 
\nonumber \\
& -\beta_{\mathrm{2}} \Delta \mu \, M' \,,
\label{eq:Dr}
\end{align}    
where the coupling coefficients are scalar in an isotropic system.
In \eqref{eq:SigmaP},
$\eta$ and $\eta'$ denote the tissue shear and bulk viscosity, respectively,
and $\mu$, $\mu'$, and $ \mu''$ 
denote their anisotropic correction depending on the cellular shape.
The term $\nu_1D'$ of \eqref{eq:Dr}
plays a role similar to that of the Gordon-Schowalter process
in the rheology of polymer melts, which describes the relaxation of polymer 
deformation, and, consequently, stress, by slippage \cite{Larson1983}.
The cross terms, including $\nu_2, $ and $\nu_3$, are possible in general. 
We introduced the active stress tensor as
$\sigma_{\mathrm{a}} =  -  \left( \zeta_1 I 
+ \zeta_2 M' \right) \Delta \mu$. 
Using the terminology of active nematic liquid crystals, 
a negative and positive
$\zeta_2$ values correspond to a contractile and extensile, respectively,
material \cite{Marchetti2013}.
These activities are often attributed to myosin contractility, for which ATP is consumed.
The terms coupling $D_{\mathrm{r}}$ and  ${\sigma_{\mathrm{e}}}'$
in \eqref{eq:Dr} underlie the Maxwellian dynamics of the system, and include
the positive coefficient $\eta_1$ with
the dimension of viscosity.
In \eqref{eq:Dr}, active cell rearrangements contribute to the constitutive 
equation for $D_{\mathrm{r}}$  as
$D_{\mathrm{a}} =  -\beta_2 \Delta \mu \, M'$.
Both $\sigma_{\mathrm{a}}$ and $D_{\mathrm{a}}$ are symmetric second-order tensors.
Our treatment of active stresses and active cell rearrangements 
was similar to that suggested previously \cite{Etournay2015,Popovic2016}, since both
approaches are inspired by the active gel models \cite{Kruse2005,Marchetti2013}.

\smallskip
Finally, the force balance equation was used to close the system:
\begin{equation}
  \label{eq:forcebalance}
\nabla \cdot \sigma  = - \vec{f}_{\textrm{ex}},
\end{equation} 
where $\vec{f}_{\textrm{ex}}$ represents the external force field,
supplemented with the appropriate boundary conditions.
Collectively, the constitutive equations 
(\ref{eq:Sigma_e},\ref{eq:totalstress},\ref{eq:SigmaP},\ref{eq:Dr})
with the kinematic relationships 
(\ref{eq:deformation_dec},\ref{eq:Kinematics})
and the force balance equation \eqref{eq:forcebalance}
determine hydrodynamic equations of a tissue (SI Sec.~2).

\section{Applications}
We investigated three simple examples of dynamical behavior predicted 
by our model, 
including the passive response following the axial stretch induced by an external force,
the deformation of a tissue due to the active internal forces, and the generation of shear flow.
Two assumptions were used for simplicity to obtain the following analytical solutions: 
2D incompressibility of a tissue ($M_0$ is constant and 
$\Tr{D} = \mathrm{div} \vec{v} = 0$; SI Sec.~2) 
and spatial homogeneity of all relevant fields.

\subsection{Passive relaxation following the axial stretching}
In \textit{Drosophila} pupal wing, an external force from the proximal part of the body 
is responsible for the stretching of the wing along the proximal-distal (PD) axis. 
Upon the tissue stretching, wing cells elongate along the PD axis, 
while the tissue relaxes during several hours when cells intercalate and
adopt a less elongated shape \cite{Aigouy2010, Sugimura2013, Guirao2015, Etournay2015}. 
Below, we demonstrated that our model can recapitulate this 
process, and determined the characteristic relaxation time 
in terms of cell mechanical parameters. 

\begin{figure}[p]
\centering
 \resizebox{0.49\textwidth}{!}{%
   \includegraphics{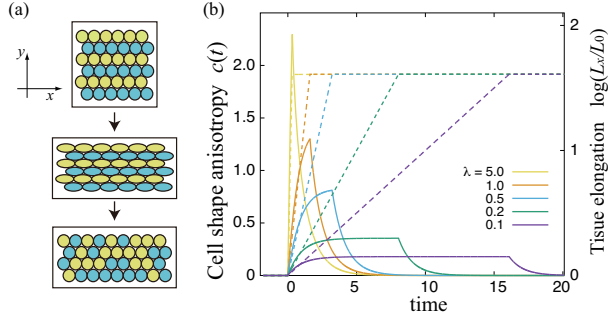}
  }
  \caption{\label{fig:FigS}
    \textbf{Passive relaxation of cellular shape anisotropy following
	the tissue stretching.}
  (a) Stretching along the $x$-axis of a tissue with the constant area. 
  Cells are elongated, and after relaxation, they recover their circular shapes.
  (b) Cellular shape anisotropy $c(t)$ (solid lines). 
  Forced deformation ($a = 5$, dashed lines) with the deformation rates 
  $\lambda  = 5.0, 1.0, 0.5, 0.2, 0.1\, \times 10^{-4} \,\mathrm{s}^{-1}$ 
  is followed by relaxation (numerical solution of \eqref{eq:cvxdr}).
  Parameters are $\gamma_0/(4M_0)^{1/2} = 0.1 \,\mathrm{mN}\,\mathrm{m}^{-1}$ , 
  $2\pi \kappa_0 = 0.4 \,\mathrm{mN}\,\mathrm{m}^{-1}$,
  $\nu_1 = 0.1$, and $\eta_1 = 1.0 \,\mathrm{Pa}\,\mathrm{m}\,\mathrm{s}$. 
  The characteristic time %is 
$\tau_{\mathrm{r}} = \eta_1/2\Gamma(0) = 1.0 \times10^4 \,\mathrm{s}$
is used as the time unit.
}
\end{figure}

We considered a tissue with an initial size $L_x \times L_y = L_0 \times L_0$ 
and in an initial isotropic, uniform state where the cell shape tensor is 
$M \!=\! M_0\,I$. From time $t=0$, the tissue elongates along the $x$-axis
at the constant rate $\dot{L}_x/L_x = \partial_x v_x = \lambda$, and
consequently, contracts along the $y$-axis at the rate 
$\dot{L}_y/L_y =\partial_y v_y = - \lambda$ (Fig.~\ref{fig:FigS}(a)).
When the tissue size reaches $a L_0 \times a^{-1} L_0$, the stretching stops
($\partial_x v_x = \partial_y v_y = 0$). 
We attempted to identify a uniform solution to the problem,
so that \eqref{eq:forcebalance} is automatically verified when 
$f_{\textrm{ex}}= 0$. All tensor variables are diagonal
($M = M_0 e^{c \Theta}$ with $\theta = 0$ ).
The $M$-dependent terms in Eqs~(\ref{eq:SigmaP}-\ref{eq:Dr}) are either canceled, 
or are isotropic tensors that may be absorbed into the pressure. 
Since we consider a passive process here and ignore active terms, 
we set $\Delta \mu = 0$. Using \eqref{eq:Kinematics}, the time 
evolution of $c$ can be written as:
\begin{equation}
  \dot{c} = 2(1-\nu_1) \partial_x v_x -2\eta^{-1}_1 \Gamma(c) \sinh(c)  \,,
\label{eq:cvxdr}
\end{equation}
where $\Gamma(c) = \gamma_0/2 \sqrt{M_0\cosh c}+ 2 \pi \kappa_0$
represents the strength of cell junction tensions as a function of
cell mechanical parameters $\gamma_0$ and $\kappa_0$.
The normal stress components were
\begin{align}
\sigma_{xx} &= -p + \eta \, \partial_x v_x + 
( 1 - \nu_1 )  \Gamma(c) \sinh c \,,\\
\sigma_{yy} &=  -p + \eta \, \partial_y v_y 
-  (1- \nu_1 )  \Gamma(c)  \sinh c \label{eq:sigyy} \,,
\end{align}
where the pressure $p$ was determined in order to satisfy the 
incompressibility condition.

Setting $a = 5$, the time course of cell shape anisotropy
$c(t)$ is presented for several values of $\lambda$ in Figure~\ref{fig:FigS}(b).
When $|c| \ll 1$,
the temporal evolution becomes Maxwellian,
\begin{equation}
  \label{eq:stretch:Maxwell}
  \dot{c} + 2 \eta^{-1}_1 \Gamma(0) \,c = 2 (1-\nu_1) \partial_x v_x.
\end{equation}
This equation explicitly relates the relaxation time 
for cell rearrangements $ \tau_{\mathrm{r}} = \eta_1/2 \Gamma(0)$
to cell mechanical parameters. 
If the stretch rate is slower than this time scale 
($\tau_{\mathrm{r}} < \lambda^{-1}$), the cells remain approximately circular 
during cell rearrangement, which is in a sharp contrast to transient 
cell elongation during the more rapid tissue stretch.

Given cell junction tensions of the order of $\gamma_0 \approx 10^{-10}
\,\mathrm{N}$ \cite{Bambardekar2015}, and a cellular length scale of the 
order of $r \approx 10^{-6}$ m, we expected
$\Gamma(0) \sim \gamma_0/r \approx 10^{-4}\, \mathrm{N}\,\mathrm{m}^{-1}$. 
Since the time scale for relaxation is of the order of a few hours, 
$\tau_{\mathrm{r}} \approx 10^4\,\mathrm{s}$ \cite{Aigouy2010,Sugimura2013}, 
we obtained the order of magnitude of the 2D viscosity coefficient 
$\eta_1 \sim \Gamma(0) \tau_{\mathrm{r}} \approx 1$ Pa m s.
For comparison, we expected a 2D shear viscosity of the order
of $\eta \approx 1 \,\mathrm{Pa}\,\mathrm{m}\,\mathrm{s}$, provided by 
$\eta = h \, \eta_{3\mathrm{D}}$, where $h \approx 10^{-5} \, \mathrm{m}$ 
represents the typical height of the epithelium and
$\eta_{3\mathrm{D}} \approx 10^5 \, \mathrm{Pa}\,\mathrm{s}$ was determined 
\emph{in vitro} in cell aggregates \cite{Guevorkian2010,Stirbat2013}.

\subsection{Active contraction-elongation (CE)}
\begin{figure}[p]
\centering
 \resizebox{0.44\textwidth}{!}{%
     \includegraphics{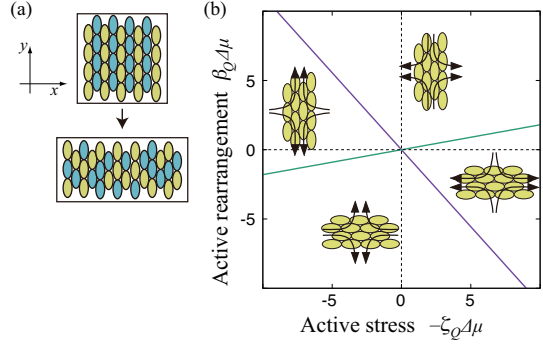}
  }
  \caption{\label{fig:FigActive}
\textbf{Active contraction-elongation (CE) of a tissue.}
  (a) Schematics of CE in \emph{Xenopus} embryo \cite{Tada2012, Shindo2014}.
  Due to the cell rearrangement, the tissue simultaneously shrinks along the 
  medio-lateral, $y$ axis and elongates along the anterior-posterior, $x$-axis,
  whereas the cells adopt an elongated shape along the $y$-axis.
  (b) Phase diagram showing the dependence of the cellular shape anisotropy $c$ 
  and the tissue deformation rate $\partial_x v_x$ on the active parameters
  $\zeta_{\mathrm{Q}} \,\Delta \mu$ and $\beta_{\mathrm{Q}} \,\Delta \mu$. 
  Ellipses illustrate   cellular shape, and arrows point to the direction of tissue flow.
  Parameter values are set as $\nu_1 = 0.1~<\!1$, 
  $\eta_1 = 1.0 \,\mathrm{Pa}\,\mathrm{m}\,\mathrm{s}$, 
  and $\eta = 5.0 \,\mathrm{Pa}\,\mathrm{m}\,\mathrm{s}$.
}
\end{figure}

CE denotes the simultaneous shrinkage and expansion 
of a tissue along two orthogonal axes \cite{Tada2012}, 
often controlled by the anisotropic localization/activity
of signaling/driving molecules, such as molecular motors 
\cite{Guillot2013, Heisenberg2013, Bosveld2012, Shindo2014, Rozbicki2015}.
During the CE, cells are often elongated along the axis perpendicular to the 
direction of the tissue flow (Fig.~\ref{fig:FigActive}(a))
\cite{Tada2012, Shindo2014, Rozbicki2015, Morishita2017}
which may occur in order to facilitate the force transmission along the axis of tissue contraction, 
through the formation of multicellular myosin cables through the 
mechanosensing of neighboring cells \cite{Rozbicki2015}.
However, the mechanisms whereby tissue deformation 
due to the cellular shape alterations counteract that due to the cell 
rearrangements remain unclear.

Here, we investigated the CE by extending our model to include 
active stress and rearrangements provided by signaling molecules
oriented along a fixed direction,
as represented by a traceless tensor
$Q = \vec{n} \otimes \vec{n} - \Tr{(\vec{n} \otimes \vec{n})}/2$, where
$\vec{n} = (\cos \phi, \sin \phi)^T$
represents a unit vector field pointing to the direction $\phi$.
Possible feedbacks on the signal activity were ignored.
We considered the lowest-order active contributions
\begin{align}
\label{eq:sigma_a}
\sigma_{\mathrm{a}} &=  -\zeta_1 \, \Delta \mu \, I 
-\zeta_{\mathrm{Q}} \, \Delta \mu \, Q \,, \\
\label{eq:D_a}
D_{\mathrm{a}} &= - \beta_{\mathrm{Q}}  \, \Delta \mu \, Q \,,
\end{align}
where the parameters $\zeta_{\mathrm{Q}}$ and $\beta_{\mathrm{Q}}$, respectively, 
quantify the strength of the active stress and of the active rearrangements.
Negative and positive $\zeta_{\mathrm{Q}}$ values correspond to 
the contractile and extensile, respectively, stress along the direction $\vec{n}$.
For positive $\beta_{\mathrm{Q}}$, $D_{\mathrm{a}}$ drives cell rearrangements 
where a cell junction parallel to $\vec{n}$ shrinks and 
is remodeled to form a new cell junction perpendicular to $\vec{n}$.
As above, $-\zeta_1 \, \Delta \mu \, I$ 
is absorbed into the pressure term when the tissue is incompressible.

We considered a uniform and fixed signal $\vec{n} = (0,1)^T$.
We set $\zeta_2 = \beta_2 = 0$ to focus on the 
activity induced by $Q$ (SI Sec.~3.1 for full calculation). 
Assuming as above that $\theta = 0$,
Eqs.~(\ref{eq:cvxdr}-\ref{eq:sigyy}) become
\begin{align}
\dot{c} &= 2(1-\nu_1) \partial_x v_x - 2\eta^{-1}_1 \Gamma(c) \sinh c  
- \beta_{\mathrm{Q}} \Delta \mu \label{eq:actiev:cvxdr}\\
\sigma_{xx} &= -p + \eta \, \partial_x v_x + 
( 1 - \nu_1 )  \Gamma(c) \sinh c + \frac{\zeta_{\mathrm{Q}}}{2} \Delta \mu \\
\sigma_{yy} &= -p + \eta \, \partial_y v_y - 
( 1 - \nu_1 )  \Gamma(c) \sinh c - \frac{\zeta_{\mathrm{Q}}}{2} \Delta \mu 
\label{eq:active:sigyy} \,.
\end{align}
In isotropic stress conditions ($\sigma_{xx} = \sigma_{yy}$), both
the cellular shape anisotropy at steady state ($\dot{c}=0$), determined by: 
\begin{equation}
\label{eq:Cdeterminanteq}
\Gamma(c) \sinh c = 
- \frac{ (1-\nu_1)\eta_1 \zeta_{\mathrm{Q}} + \eta \eta_1 \beta_{\mathrm{Q}}  }{ \eta_1(1-\nu_1)^2 + \eta}
\frac{\Delta \mu}{2} \,,
\end{equation}
and the tissue deformation rate $\partial_x v_x$ 
\begin{equation}
\label{eq:Vx}
\partial_x v_x =
 \frac{ -\zeta_{\mathrm{Q}} +   (1-\nu_1) \eta_1\beta_{\mathrm{Q}}}{ \eta_1 \left(1-\nu_1\right)^2 + \eta } \frac{\Delta \mu}{2}  
   \,.
\end{equation}
remain non-zero, indicating that the tissue anisotropy and
flow are maintained through the active stresses and cell rearrangements.
In Eqs.~(\ref{eq:Cdeterminanteq}) and  (\ref{eq:Vx}), cellular shape anisotropy
$c$ and velocity gradient $\partial_x v_x$ may adopt either an identical or
an opposite sign depending on the numerical values of the active coefficients 
$\beta_{\mathrm{Q}}$ and $\zeta_{\mathrm{Q}}$.
Therefore, cell elongation occurs either parallel or perpendicular 
to the direction of flow, depending on the parameter values
(phase diagram in Fig.~\ref{fig:FigActive}(b), 
with $\nu_1 < 1$). 
Considering the contractile effect of myosin motors
(upper right quadrant of Fig.~\ref{fig:FigActive}(b),
$- \zeta_{\mathrm{Q}}>0$ and $\beta_{\mathrm{Q}} > 0$),
cell elongation occurs mostly perpendicular to the tissue flow, 
except below the green line of the slope $(1-\nu_1)/\eta$.
An earlier study using CPM suggested that the differential cell adhesion accounts 
for CE with cell elongation orthogonal to tissue flow, 
in which only the outer tissue boundary contributes to the driving of 
tissue deformation \cite{Zajac2000,Zajac2003}.
Our model provides an alternative mechanism in which activities play
an essential role, in agreement with recent observations of elevated myosin activity in the elongated cell junctions orthogonal to the tissue flow \cite{Shindo2014, Rozbicki2015}.

%%%%% Applications 3%%%%%%%
\subsection{Shear flow}
\begin{figure*}[bt]
\centering
 \resizebox{0.65\textwidth}{!}{%
   \includegraphics{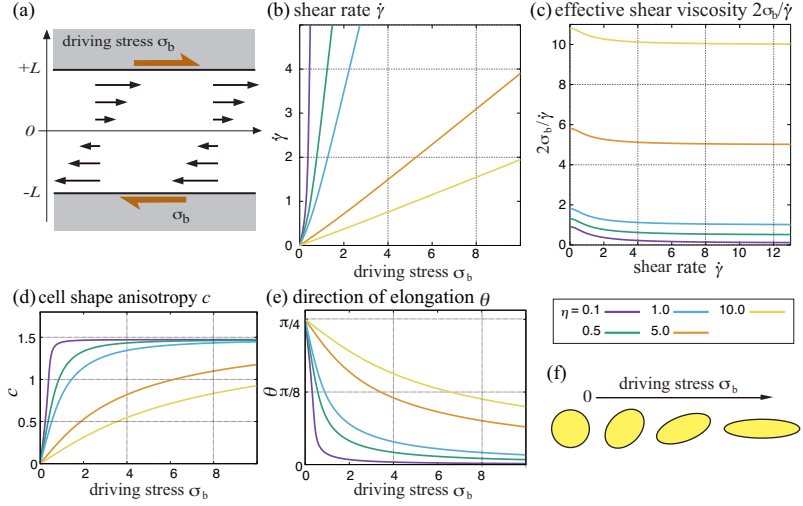}
  }
  \caption{\label{fig:FigShear}
  \textbf{The tissue behaving as a shear-thinning fluid.}
(a) An external shear stress
($\sigma_{\mathrm{b}}$) is applied at the top and 
bottom boundaries ($y = \pm L$).
(b) The shear rate $\dot{\gamma}$ is plotted against 
the driving stress $\sigma_{\mathrm{b}}$.
(c) The effective viscosity $2\sigma_{\mathrm{b}}/\dot{\gamma}$ 
decreases as a function of the shear rate $\dot{\gamma}$.
Cell shape anisotropy $c$ (d) and orientation $\theta$ (e) are plotted as
a function of the driving stress $\sigma_{\mathrm{b}}$. 
(f) Cells elongate and align along the force axis as the
driving stress $\sigma_{\mathrm{b}}$ increases. Parameters were set as 
$\gamma_0/(4M_0)^{1/2} = 0.1 \,\mathrm{mN}\,\mathrm{m}^{-1}$, 
$2\pi\kappa_0 = 0.4 \,\mathrm{mN}\,\mathrm{m}^{-1}$,  $\nu_1 = 0.1$ and
$\eta_1= 1.0 \,\mathrm{Pa}\,\mathrm{m}\,\mathrm{s}$.
}
\end{figure*}

%%%%%%%%%%%%
A fundamental geometry for investigating rheology \cite{Oswald2009},
shear flow is commonly found in many developmental 
tissues \cite{Guirao2015,Sato2015,Rozbicki2015}.
Here, we considered the simple geometry
given in Fig.~\ref{fig:FigShear}(a), inspired by the plane Couette flow \cite{Oswald2009}.
The flow, with shear rate $\dot{\gamma} = \partial_y v_x$, is driven by an external
shear stress $\sigma_{\mathrm{b}}$ acting in the opposite directions on the boundaries. 
The effective shear viscosity $\eta_{\mathrm{eff}}$ can be calculated as:
\begin{equation}
\eta_{\mathrm{eff}} = \frac{2\sigma_{\mathrm{b}}}{\dot{\gamma}} = 
\eta + (1-\nu_1)^2 \eta_1 -(1-\nu_1)\eta_1\cos 2\theta \tanh c \,,
\end{equation}
where the cell shape anisotropy $c$ and orientation $\theta$
depend on the driving stress $\sigma_{\mathrm{b}}$ (SI Sec.~3.2).
In the presence of a coupling between the cellular rearrangement rate and
the elastic stress ($\eta_1 \ne 0$), $\eta_{\mathrm{eff}}$ depends on 
$\sigma_{\mathrm{b}}$, which makes the tissue non-Newtonian 
(Fig.~\ref{fig:FigShear}). 
The shear rate is an increasing function of the external stress
$\sigma_{\mathrm{b}}$,
whereas the effective shear viscosity decreases with $\dot{\gamma}$, 
indicating that the model predicts shear-thinning 
(Fig.~\ref{fig:FigShear}(b-c)). Cellular shape anisotropy $c$ increases 
with $\sigma_{\mathrm{b}}$ or $\dot{\gamma}$, to converge to a finite value 
for large driving. Cells turn in the direction of the applied stress
as they elongate (Fig.~\ref{fig:FigShear}(d-f)).
Shear-thinning was reported \emph{in vitro}, using cellular spheroids
\cite{Marmottant2009}, and was shown to be related to stress-dependent barriers that 
may control cell rearrangements (see \cite{Marchetti2013} for
a mechanism leading to shear-thinning in active materials that does
not involve topological effects). To the best of our knowledge, there is no experimental evidence for the shear-thinning in epithelial tissues \emph{in vivo}, which is a non-trivial 
prediction of our model, obtained assuming only linear force-flux couplings.

Including the active stress and active cell rearrangements, both 
internal, Eqs.~(\ref{eq:SigmaP}-\ref{eq:Dr}), or due to an oriented signal, 
Eqs.~(\ref{eq:sigma_a}-\ref{eq:D_a}), the shear rate becomes
\begin{equation}
\dot{\gamma} = 2 \, \frac{\sigma_{\mathrm{b}} + \sigma_2  + \sigma_{\mathrm{Q}}}
{\eta + (1-\nu_1)^2 \eta_1 -(1-\nu_1)\eta_1\cos 2\theta \tanh c} \,,
\end{equation}
with $\sigma_2 = \Delta \mu \, M_0\,\sinh c \, \sin 2 \theta  
( \zeta_2- 2 \eta_1 (1-\nu_1) \beta_2)$ 
and $\sigma_{\mathrm{Q}} = \Delta \mu \,\sin 2 \phi 
( \zeta_{\mathrm{Q}}+ \eta_1 (1-\nu_1) \beta_{\mathrm{Q}})$.
In addition to the external stress
$\sigma_{\mathrm{b}}$, active stresses and 
active cell rearrangements are able to drive shear flow. 
Indeed, the active rearrangements described by \eqref{eq:D_a}
produce shear flow for an arbitrary orientation $\phi$, as has been 
observed in the genitalia of \emph{Drosophila} and demonstrated using the CVM 
in the case of $\vec{n}$ pointing to $\phi = 3\pi/4$ 
\cite{Sato2015}.

%%%%% Discussion %%%%%%%
\section{Discussion}
We formulated a continuum model of epithelial mechanics based on
the definition of a tissue-scale field $M$ representing cellular shape.
The advantages of using this approach are as follows. 
Most importantly, our model is designed to connect 
cellular level mechanical ingredients (\emph{e.g.}, cell area elasticity 
and cell junction tension) and cell morphogenetic processes (\emph{e.g.}, cell rearrangements),
in order to drive tissue mechanics and deformation.
This was achieved by defining the energy function and kinematic relationship 
in terms of the cell shape field $M$, which distinguishes our work
from the previous continuum models \cite{Tlili2015,Popovic2016,Brodland2006}.
The model describes time-dependent flows, and allows the evaluation of time scales as a function of material parameters. 
Large and non-affine deformations can be treated.
In addition, the model  can also incorporate a signal field, for instance,
the axial tensor $Q$, which, here, orients active stresses and cellular rearrangements.
Many relevant fields can be experimentally determined, 
including tissue stress, tissue deformation, cell morphogenetic processes, 
and chemical signaling fields, such as the concentrations of cell polarity 
molecules or the orientation field describing the spatial distribution of myosin molecules.
Once their dynamics are quantitatively characterized by the relevant scalar, vector, or tensor variables
\cite{Aigouy2010, Sugimura2013, Bosveld2012, Guirao2015, Ishihara2012, Blanchard2009, Etournay2015, Blanchard2017},
comparison between the model and experiments 
is feasible.

The data presented here demonstrate that using our model, we can predict the dynamical 
behaviors underlying epithelial tissue morphogenesis.
In the future, the quantitative comparison of the model predictions with experimental data should
help us evaluate material parameters and validate constitutive equations.
For example, since $D$, $D_{\mathrm{r}}$ \cite{Guirao2015} and 
$\sigma_{\mathrm{e}}$ \cite{Ishihara2012, Ishihara2013} are 
measurable quantities, the validity of \eqref{eq:Drc} can be tested by comparing it with the experimental data.
Furthermore, the relevance of the non-linear terms in \eqref{eq:Dr} for epithelial rheology may be tested.

The current approach can be extended in several ways.
Other cell morphogenetic processes, such as cell division and cell death, should be incorporated to the current modeling scheme \cite{Yabunaka2017}. 
Plastic behavior \cite{Marmottant2009} may be considered as well, either
within the dissipation function formalism \cite{Tlili2015}, or
by considering non-linear constitutive equations to 
effectively incorporate a yield stress. 
In analogy to the recent adaptations of the CPM \cite{Kabla2012} 
or the CVM \cite{Barton2016}, a cell polarity field may be included to describe collective cell migration \cite{Yabunaka2017}. 
Another possible extension of the model 
concerns kinetics. 
Here, the associated dissipation coefficients, %as well as 
including the coefficients governing active stress and active rearrangement, were determined phenomenologically by employing the thermodynamic formalism. 
This point can be further explored by considering detailed processes at the cellular level. 
In particular, the cell-level machinery
underlying tissue-level active processes should be studied further in connection with 
signal activity dynamics. 
Finally, our 2D formalism can be extended to 3D.

In conclusion, the present work provides an integrated scheme
for the understanding of the mechanical control of epithelial morphogenesis. 
Dynamics of signal fields can be coupled to the equations.
Feedback between biochemical signaling and mechanics through the mechanosensing of a cell
\cite{Blanchard2011, Aigouy2010, Sugimura2013, Aw2016, Uyttewaal2012} may represent a potential future research direction.

\paragraph{acknowledgements}

We thank Yohanns Bella\"iche, 
Cyprien Gay, Fran\c{c}ois Graner, Boris Guirao,  
and Shunsuke Yabunaka for discussion. 
This research was supported by JSPS KAKENHI Grant Number JP24657145 and JP25103008, 
the JST CREST (JPMJCR13W4), the JST PRESTO program (13416135), and by the JSPS/MAEDI/MENESR Sakura program.

%\showacknow{} % Display the acknowledgments section

% \pnasbreak splits and balances the columns before the references.
% Uncomment \pnasbreak to view the references in the PNAS-style
% If you see unexpected formatting errors, try commenting out \pnasbreak
% as it can run into problems with floats and footnotes on the final page.
% \pnasbreak

% Bibliography
% \bibliography{pnas-sample}
% implement with bibtex file if possible

%%%%% References %%%%%

\break

\appendix

\section{Energy function and elastic stress}

\subsection{Derivation of elastic stress}
The shape of a given cell is represented by 
$(\vec{r}-\vec{r}_c)^T \, M^{-1} \, (\vec{r}-\vec{r}_c) = 1$, 
where $M$ is a positive definite matrix, and
the center of the cell is located at $\vec{r}_c$. 

Let each material point at the position $\vec{r}$ move to 
$\vec{r'} = \vec{r} +\vec{u}(\vec{r})$, thus defining the displacement 
field $\vec{u}$.
The center of a cell changes as $\vec{r'}_c = \vec{r}_c+ \vec{u}(\vec{r}_c)$
and the cell shape changes as
\[
(\vec{r}-\vec{r'}_c)^T  \left( (1+\nabla\!\vec{u})^{-1}\right)^T
M^{-1}(1+\nabla\!\vec{u})^{-1}(\vec{r}-\vec{r'}_c) = 1 \,.
\]
Upon coarse-graining, this indicates that $M$ 
changes as
\begin{equation*}
  M'(\vec{x}+\vec{u}) = (1+\nabla \vec{u}) M(\vec{x}) (1+\nabla \vec{u}^T)\,,
\end{equation*}
whereby, at order $\mathcal{O}(|\nabla \vec{u}|)$, 
\begin{equation}
  M \to M' = M -\vec{u}\cdot \nabla M + M (\nabla\! \vec{u})^T + \nabla \vec{u} M
\,.
 \label{eq:Mchange}
\end{equation}
This equation represents the relationship between the change in the cell shape 
and tissue displacement field, in the absence of
cell rearrangements.
It has been derived rigorously for a cellular material in \cite{Tlili2015}.

By the virtual displacement
$\vec{u}$, the total energy  $\mathcal{F}(M) = \int F(M) \mathrm{d}\vec{x}$
changes as follows:
\begin{align}
  \delta \mathcal{F} &= \mathcal{F}(M') - \mathcal{F}(M) \simeq 
  \int \frac{\partial F}{\partial M} :\delta M \mathrm{d}\vec{x} \nonumber\\
  &=
 \int \frac{\partial F}{\partial M} :\left( -\vec{u}\cdot \nabla M +M (\nabla\! \vec{u})^T + \nabla \vec{u} M\right) \mathrm{d}\vec{x}
  \nonumber \\
  &=\int \left[-\nabla \!\cdot\! (\vec{u}F)+
 \left(FI+ \left(\frac{\partial F}{\partial M} \right)^TM+\frac{\partial F}{\partial M}M^T\right): \nabla \vec{u} \right]\mathrm{d}\vec{x} \nonumber
 \end{align}
where $I$ represents the unit tensor and $\delta M = M'-M$.
The first term vanishes at the boundary of the system, and
the elastic stress $\sigma_{\mathrm{e}}$ is given as:
\begin{equation}
  \sigma_{\mathrm{e}} = FI+\left(\frac{\partial F}{\partial M}\right)^TM + \left(\frac{\partial F}{\partial M}\right)M^T \,,
\end{equation}
Since $M$ is symmetric, this expression further simplifies to
\begin{equation}
\label{eq:def:sige}
  \sigma_{\mathrm{e}} = FI+ 2 \left(\frac{\partial F}{\partial M}\right) M \,.
\end{equation}

Using the relations
\[  \DP{}{M} \Tr{M} = I, \quad \DP{}{M}|M| = |M| M^{-1} \,, \]
and given Eq.~(7) for the expression of the energy,
Eq.~\eqref{eq:def:sige} leads to
\begin{equation}
\label{eq:Sigma_e}
\sigma_{\mathrm{e}} =    K(\pi |M|^{1/2} -A_0)I + 
\left(
\frac{\gamma_0}{\sqrt{2 \Tr{M}}}
+  2 \pi \kappa_0 
\right) \, \frac{M}{|M|^{1/2}}  \,.
\end{equation}

\subsection{Comparison of macroscopic and microscopic stress} 
To check the validity of coarsening, we conducted numerical simulation of CVM 
with an energy function given by Eq.~(3) in the main text 
\cite{Farhadifar2010,Fletcher2014}:
\begin{equation}
f_{c} =  \sum_i \frac{K}{2}\left(A_i-A_0\right)^2 + \sum_{[ij]} \gamma_0 \ell_{ij} + \sum_i\frac{\kappa_0}{2}L_i^2 \,,
\label{eq:CVM_energy}
\end{equation} 
and compared two expressions of the stress tensor.

The first one is the `microscopic' expression directly calculated from CVM \cite{Ishihara2012, Ishihara2013, Guirao2015}
\begin{equation}
\label{eq:CVMStress}
\sigma^{\mathrm{CVM}}_{} = \frac{1}{\sum_i A_i} \left( - \sum_i  P_i A_i I+ \sum_{[ij]} T_{ij} \frac{\ell_{ij} \otimes \ell_{ij}}{|\ell_{ij}|} \right) \,
\end{equation}
where $A_i$ is the area of cell $i$, and $\ell_{ij}$ is the length of the interface between cells $i$ and $j$.
$P_i$ is the pressure of $i$-th cell originating from cell elasticity, 
and $T_{ij}$ is the tension of cell interface $[ij]$.
$P_i$ and $T_{ij}$ are determined from the above energy function
as $P_i = -K(A_i -A_0)$ and 
$T_{ij} =\gamma_{\mathrm{0}} + \kappa_{\mathrm{0}} \left( L_i + L_j\right)$.

The second one is the `macroscopic' expression of stress %expressed 
obtained by coarse-grained cell shape $M$, Eq.~\eqref{eq:Sigma_e},
estimated from the given geometry of the cell in a CVM simulation.
In practice, we calculated the centroid and the
second moment $\mu^i_2$ of each polygon (cell) $i$ \cite{Steger1996}, and 
then averaged it over $N$ cells:
\begin{equation}
 M = \frac{4}{N}\sum_i{\mu^i_2} = \frac{4}{N}\sum_i
 \begin{pmatrix}
  \mu^{i}_{2,xx} &   \mu^{i}_{2,xy} \\
   \mu^{i}_{2,yx} &   \mu^{i}_{2,yy}
 \end{pmatrix}
\end{equation}
The factor $4$ is needed since the second moment of an ellipse with 
major and minor radii $a$ and $b$ has eigenvalues $a^2/4$ and $b^2/4$. 
With the estimated $M$, we calculated the cell area as $\pi|M|^{1/2}$,
and the cell perimeter by using Euler's formula for the ellipse perimeter 
to the second order \cite{Chandrupatla2010}:
\begin{equation}
 L(M) = c_{h} \pi \sqrt{2\Tr{M}} \left[ 1 - \frac{1}{16} \left( 1-\frac{4|M|}{(\Tr{M})^2}\right)\right]
\end{equation}
Taking into accout the factor $c_{h} = \sqrt{2\sqrt{3}/\pi} \sim 1.05$, the 
ratio between the perimeters of a circle and an hexagon with the same area,
slightly improves the fitting.
The precision of Euler's expansion for the ellipse perimeter is 
illustrated in Fig.~S1.

CVM simulations were conducted by minimising the energy \eqref{eq:CVM_energy}.
An external stress $\sigma^{\mathrm{ex}}$ was applied on the boundary of the system,
for which we took $\sigma^{\mathrm{ex}}_{xy} = \sigma^{\mathrm{ex}}_{yx} 
= \sigma^{\mathrm{ex}}_{yy} = 0$, while $\sigma^{\mathrm{ex}}_{xx}$ was 
controlled to stretch the system along the $x$-axis. After the system 
relaxed, we confirmed that the force 
was balanced and that the stress $\sigma^{\mathrm{CVM}}$ converged to 
coincide with the external stress $\sigma^{\mathrm{ex}}$. 
In Fig.~S2, we distinguish the stress that comes from cell elasticity 
($- P^{\mathrm{ce}} \, I$, 
where $P^{\mathrm{ce}}$ denotes the pressure) 
and from cell junction tension ($\sigma^{\mathrm{T}}$),
respectively (i.e., the first and the second 
terms of Eqs.~\eqref{eq:CVMStress} and \eqref{eq:Sigma_e}). 
In the simulations, parameters are set as $K = 10.0$, $A_0 = 1.0$ and 
$\kappa_0/KA_0 = 0.02$, $0.04$.
The results are summarized in Fig.~2(a,b) in the main text
and detailed in Fig.~S2. 

The values found for the macroscopic expression of stress with coarse-grained 
cell shape $M$ agree well with the microscopic (correct) stress, as long as
the cell aspect ratio is not too large.

%\clearpage

%%%%%%Stability analysis%%%%%%%
\subsection{Stability analysis of the energy function} 

\paragraph{Cell area instability}
Let cell shape be uniform, not depending on the position $\vec{r}$. 
With the expression $M = M_0 e^{c\Theta}$,
the cell area is taken as $A = \pi M_0$, and the energy 
density function per cell, $f = A \,F(M)$, is expressed as
\begin{equation}
f = \frac{K}{2}\left(\pi  M_0-A_0\right)^2 + 
\pi  \gamma_0 \sqrt{M_0 \cosh c}
+2\pi^2\kappa_0 M_0 \cosh c
\label{eq:fm_S0C}
\end{equation}
The pressure $P$ is obtained by differentiating $f$ with respect to 
$A = \pi M_0$ (equivalently $P = -\Tr{\sigma_{\mathrm{e}}}/2$)
\begin{equation}
 P = - \frac{\partial f}{\partial (\pi M_0)}
  = -K(\pi  M_0-A_0) 
  -\frac{\gamma_0}{2} \sqrt{\frac{\cosh c}{M_0}}
   -2 \pi \kappa_0  \cosh c
\label{eq:Pressure}
\end{equation}
Thermodynamic stability holds when $\partial P/\partial A < 0$, which leads to the following condition:
\begin{equation}
\overline{\gamma}_0 \equiv \frac{\gamma_0}{KA^{3/2}}<  \frac{4}{\pi^{1/2}\cosh c}
\end{equation}
Note that $\cosh c \geq 1$, where the equality holds at $c= 0$. 
The cell area $A$ depends on parameters and boundary conditions.
For a given cell area, the described condition does not hold for large $\gamma_0$ values, indicating that the homogenous
cell size state becomes unstable. 
Taking higher order in approximating ellipse perimeter does not change the condition.

\paragraph{Cell shape instability}
$f$ is an even function with respect to $c$, and is expanded as $f(c) = f(0) + \frac{f''(0)}{2!}c^2 + \frac{f^{(4)}(0)}{4!}c^4+\cdots $, where
\begin{equation}
f''(0) =  \pi \frac{\gamma_0}{2} \sqrt{M_0} +2\pi^2 \kappa_0M_0  \, .
\end{equation}
$f$ takes its minimal value at $c=0$ as long as $f''(0) > 0$ 
\emph{i.e.},
\begin{equation}
  \gamma_0 > -4\kappa_0\pi M_0^{1/2} \,. \label{eq:SoftCond}
\end{equation}
If $\gamma_0$
is smaller than the threshold value
$-4\kappa_0 \pi M_0^{1/2} $, the 
circular shape is no longer
stable, and cells preferentially take an elongated shape.
Using a non-dimensionalized parameter $\overline{\kappa}_0 = \kappa_0/KA$, the above condition can be written as %(using $\hat{\pi} = \pi$ again)
\begin{equation}
  \overline{\gamma}_0 > -4\pi^{1/2} \overline{\kappa}_0
\end{equation}
This condition is unchanged when higher orderer correction of ellipse perimeter is taken into account.

%\break

\vspace{20mm}
\section{Hydrodynamic equations of epithelial mechanics}
%for tissue mechanics}

For convenience, we list the equations describing tissue hydrodynamics
derived in the main text.

\paragraph{Force balance equation} reads:
\begin{equation}
\nabla \cdot \sigma = -\vec{f}_{\mathrm{ex}} \,,
\end{equation} 
where $\vec{f}_{\mathrm{ex}}$ represents the external force.

\paragraph{Kinematic equations} are given as: 
\begin{align}
\nabla \vec{v} &= \Omega + D_{\mathrm{s}} +D_{\mathrm{r}}\\
 \dot{M} &= \left(\nabla\!\vec{v} -D_{\mathrm{r}}\right) M + M \left( \nabla\!\vec{v}-D_{\mathrm{r}} \right)^T \label{eq:Kinematics}\\  
\Tr{D_{\mathrm{r}}} &= 0 \label{eq:Kinematics:TrDr}
\end{align}
Here, $\Omega = (\nabla \vec{v}-[\nabla \vec{v}]^T)/2$ represents the asymmetric part of the deformation rate tensor,
while $D =  D_{\mathrm{s}} +D_{\mathrm{r}}$ represents its 
symmetric part. Without cell division and death,
$D_{\mathrm{r}}$ represents the deformation rate caused by cell
rearrangement, the trace of which is zero.

\paragraph{Cell number balance equation}
Given Eq.~\eqref{eq:Kinematics}, we calculate
\begin{equation}
  \label{eq:detMdot}
  \DT{}{t}|M| = |M| \, \mathrm{Tr}\left( M^{-1} \DT{M}{t} \right) 
 =  2|M| \, \left( \nabla \cdot \vec{v} - \Tr{D_{\mathrm{r}}} \right) \,.
\end{equation}
Here, $d|M|/dt$ is Lagrange derivative, \emph{i.e.,} $ \mathrm{d}|M|/\mathrm{d}t \equiv \partial |M|/\partial t + \vec{v} \nabla |M|$.
The cell number density field $\rho$ is defined by 
$\rho  = 1/\pi |M|^{1/2}$, and its evolution equation reads
\begin{equation}
  \label{eq:rhodot}
  \DP{\rho}{t} + \nabla \cdot \left(\rho \vec{v} \right) = [\Tr{D_{\mathrm{r}}}] \rho  \,.
\end{equation}
where $\Tr{D_{\mathrm{r}}}$ indicates the cell number variation rate.
Without cell division and death, cell number density does not change through 
cell rearrangement and 
$\Tr{D_\mathrm{r}} = 0$ \eqref{eq:Kinematics:TrDr}.

\paragraph{Constitutive equations} are given as:
\begin{align}
\sigma &= \sigma_{\mathrm{e}} + \sigma_{\mathrm{p}} \\
 \sigma_{\mathrm{e}} &=  K \left( \pi |M|^{1/2}-A_0\right)  I + 
\left(\frac{\gamma_0}{\sqrt{2 \Tr{M}}}
+  2 \pi \kappa_0 \right) \, \frac{M}{|M|^{1/2}} \\
\sigma_{\mathrm{p}} &=
\eta \, D' + \eta' \, \Tr{D} 
%+ \mu \, (D M'+ M' D -\Tr{(DM')}I)  
+ \mu \, (D M'+ M' D )
+ \mu' \, (\Tr{D}) M'
+\mu'' \,(\Tr{(DM')}) I
\nonumber\\
& -\nu_1 {\sigma_{\mathrm{e}}}' -\nu_2 \left( {\sigma_{\mathrm{e}}}' M' + M' {\sigma_{\mathrm{e}}}'  \right)
-\nu_3 ({\sigma_{\mathrm{e}}}'\!:\!M')I  \nonumber\\
& - \zeta_1 \Delta \mu \, I - \zeta_2 \Delta \mu \, M'
\label{eq:SigmaP}\\
D_{\mathrm{r}} &= \nu_1 D' + \nu_2 (DM'+M'D-\Tr{(DM')}I) + \nu_3 (\Tr{D}) M' \nonumber \\
&+ \eta^{-1}_1 {\sigma_{\mathrm{e}} }' +\eta^{-1}_2 \left( {\sigma_{\mathrm{e}}}'M' + M'{\sigma_{\mathrm{e}}}'- \Tr{({\sigma_{\mathrm{e}}}'M')}I \right) 
-\beta_{\mathrm{2}} \Delta \mu \, M' \,,
\label{eq:Dr}
\end{align} 
where $\sigma_{\mathrm{e}}$ and $\sigma_{\mathrm{p}}$ are the elastic and 
the dissipative stress respectively, and $\eta$ and $\eta'$ are the tissue shear and bulk viscosity.

To obtain Eqs.~(\ref{eq:SigmaP}-\ref{eq:Dr}) (Eqs.~(15) and (16) in the main text),
we set the fourth-order tensors $\chi^{\mathrm{sr}}$ and $\chi^{\mathrm{rr}}$ as follows:  
\begin{align}  
\chi^{\mathrm{ss}}_{ijkl} &= \eta \delta_{ik}\delta_{jl} + \eta' \delta_{ij}\delta_{kl}  \nonumber \\&
%+ \mu\!\left( \delta_{ik}M'_{lj} + M'_{ik}\delta_{jl}- M'_{lk}\delta_{ij}\right)+ \mu' M'_{ij}\delta_{kl}\\
+ \mu\!\left( \delta_{ik}M'_{lj} + M'_{ik}\delta_{jl}\right)+ \mu' M'_{ij}\delta_{kl} + \mu'' M'_{lk}\delta_{ij}\\
\chi^{\mathrm{sr}}_{ijkl} &=  \nu_1  \!\!\left(\! \delta_{ik}\delta_{lj}-\frac{1}{2}\delta_{ij}\delta_{kl} \!\right)  
+ \nu_2 \!\left( \delta_{ik}M'_{lj} + M'_{ik}\delta_{jl}\!-\!M'_{ij}\delta_{kl}\right)\!  \nonumber \\&+\nu_3 M'_{kl}\delta_{ij} \\
\chi^{\mathrm{rr}}_{ijkl} &= \eta_1^{-1} \delta_{ik}\delta_{jl} + \eta_2^{-1}\left(\delta_{ik}M'_{lj} + M'_{ik}\delta_{jl}- M'_{lk}\delta_{ij} \right) 
\end{align}
Here $\delta_{ij}$ is the Kronecker tensor.
$\chi^{\mathrm{rs}}_{ijkl}$ is determined by Onsager's reciprocity $\chi^{\mathrm{rs}}_{ijkl} =\chi^{\mathrm{sr}}_{klij}$:
\begin{align}
\chi^{\mathrm{rs}}_{ijkl} & = \nu_1 \!\left( \delta_{ik}\delta_{lj}-\frac{1}{2}\delta_{ij}\delta_{kl}\right) 
+ \nu_2 \left( \delta_{ik}M'_{lj} + M'_{ik}\delta_{jl}- \delta_{ij}M'_{kl}\right) \nonumber\\
&+\nu_3 \delta_{kl}M'_{ij} \,.
\end{align}
%\begin{align}  
%\chi^{\mathrm{ss}}_{ijkl} &= \eta \delta_{ik}\delta_{jl} + \eta' \delta_{ij}\delta_{kl}\\
%\chi^{\mathrm{sr}}_{ijkl} &=  \nu_1 \!\left( \delta_{ik}\delta_{lj}-\frac{1}{2}\delta_{ij}\delta_{kl}\right) + \nu_2 \left( \delta_{ik}M'_{lj} + M'_{ik}\delta_{jl}- M'_{ij}\delta_{kl}\right) %\nonumber\\
%&+\nu_3 M'_{kl}\delta_{ij}\\
%\chi^{\mathrm{rr}}_{ijkl} &= \eta_1^{-1} \delta_{ik}\delta_{jl} + \eta_2^{-1}\left(\delta_{ik}M'_{lj} + M'_{ik}\delta_{jl}- M'_{ij}\delta_{kl} \right)
%\end{align}
%$\chi^{\mathrm{ss}}$ is taken to be zero, and $\chi^{\mathrm{rs}}_{ijkl}$ is determined by Onsager's reciprocity $\chi^{\mathrm{rs}}_{ijkl} =\chi^{\mathrm{sr}}_{klij}$.
%\begin{align}
%\chi^{\mathrm{rs}}_{ijkl} & = \nu_1 \!\left( \delta_{ik}\delta_{lj}-\frac{1}{2}\delta_{ij}\delta_{kl}\right) + \nu_2 \left( \delta_{ik}M'_{lj} + M'_{ik}\delta_{jl}- \delta_{ij}M'_{kl}\right) %\nonumber\\
%&+\nu_3 \delta_{kl}M'_{ij}
%\end{align}

\paragraph{Incompressible case}

An incompressible flow is characterized by a constant $|M|$, and thus 
$\nabla \cdot \vec{v} = 0$ according to Eq.~\eqref{eq:detMdot}.
The factorization $M = M_0 \, e^{c\Theta}$ is all the more useful since 
$M_0$ is constant.

The constitutive equations are replaced by
 \begin{align}
 \sigma &= {\sigma'_{\mathrm{e}}} +{\sigma_{\mathrm{p}}}' - pI \,,\\
 {\sigma_{\mathrm{e}}}' &=
 \left( \frac{\gamma_0}{\sqrt{2 \Tr{M}}}
+  2 \pi \kappa_0 
\right) \, \frac{M'}{|M|^{1/2}}\,,\\
{\sigma_{\mathrm{p}}}' &=
\eta \, D' + \mu \, (D M'+ M' D -\Tr{(DM')}I )
\nonumber \\
&-\nu_1 {\sigma_{\mathrm{e}}}' -\nu_2 \left( {\sigma_{\mathrm{e}}}' M' + M' {\sigma_{\mathrm{e}}}'  \right)
- \zeta_2 \Delta \mu \, M'\\
 D_{\mathrm{r}} &= \nu_1 D + \nu_2 (DM'+M'D-\Tr{(DM')}I) \,, 
\nonumber \\
&+ \eta^{-1}_1 {\sigma_{\mathrm{e}} }' +\eta^{-1}_2 \left( {\sigma_{\mathrm{e}}}'M' + M'{\sigma_{\mathrm{e}}}'- \Tr{({\sigma_{\mathrm{e}}}'M')}I \right) 
\nonumber\\
& -\beta_{\mathrm{2}} \Delta \mu \, M' \,,
\end{align}
where $p$ represents the tissue pressure.

\vspace{20mm}
\section{Applications} % Case studies}
\subsection{Active contraction-elongation}
In the main text, we considered the case $\zeta_2 = \beta_2 = 0$. 
Here we take into account these terms and consider their role. 
For active contraction-elongation with signal along $\vec{n} = (0,1)$, 
the governing equations become
\begin{align}
\dot{c} &= 2(1-\nu_1) \partial_x v_x - 2\eta^{-1}_1 \Gamma(c) \sinh c  
	+ 2 \beta_2 \Delta \mu M_0\sinh c
	- \beta_{\mathrm{Q}} \Delta \mu \label{eq:actiev:cvxdr}\\
\sigma_{xx} &= -p + \eta \, \partial_x v_x + 
	( 1 - \nu_1 )  \Gamma(c) \sinh c
	-\zeta_2 \Delta \mu M_0 \sinh c
	+ \frac{\zeta_{\mathrm{Q}}}{2} \Delta \mu \\
\sigma_{yy} &= -p + \eta \, \partial_y v_y - ( 1 - \nu_1 )  \Gamma(c) \sinh c
	+\zeta_2 \Delta \mu M_0 \sinh c
	- \frac{\zeta_{\mathrm{Q}}}{2} \Delta \mu \,
\label{eq:active:sigyy} 
\end{align}
for which $\Gamma(c) = \gamma_0/2\sqrt{M_0\cosh c} + 2\pi \kappa_0$ and 
the incompressibility condition 
$\partial_x v_x + \partial_y v_y = 0$ is taken into account. 
With isotropic boundary condition $\sigma_{xx} = \sigma_{yy}$, 
$\partial_x v_x$ is given by
\begin{equation}
\eta \partial_x v_x = -(1-\nu_1)\Gamma(c) \sinh c + \zeta_2 \Delta \mu M_0 \sinh c
-\frac{\zeta_{\mathrm{Q}}}{2}\Delta \mu \,
\end{equation}
At steady state ($\dot{c} = 0$), cell shape anisotropy $c$ is determined by
\begin{equation}
\label{eq:Ceq}
\Gamma(c) \sinh c =  \frac{\eta_1\left[ - (1-\nu_1) \zeta_{\mathrm{Q}} - \eta \beta_{\mathrm{Q}}
+2\left( (1-\nu_1) \zeta_2 + \eta \beta_2 \right)  M_0 \sinh c
\right]}{\eta_1(1-\nu_1)^2 +\eta} \frac{ \Delta \mu}{2}\,.
\end{equation}
The tissue exhibits steady flow, with $\partial_x v_x $ given by
\begin{equation}
\partial_x v_x 
   =
   \frac{  -\zeta_{\mathrm{Q}} +   (1-\nu_1)\beta_{\mathrm{Q}} \eta_1+ 2 (\zeta _2 - (1-\nu_1)\beta_2\eta_1)M_0 \sinh (c))}{ \eta _1 \left(1-\nu
   _1\right)^2+\eta } \frac{\Delta \mu}{2}
\end{equation}

% SI   I removed the following sentences, because the situation seeems a bit complicated   
%In \eqref{eq:Ceq}, even without signal-associated activity $(\zeta_{\mathrm{Q}} = \beta_{\mathrm{Q}}= 0)$,
%$c$ can be non-zero for appropriate $\zeta_2$ and $\beta_2$,
%indicating spontaneous
%symmetry breaking by which cell deformation ($c \neq 0$) occurs 
%in the absence of external directional signal.

\subsection{Shear flow}
We will consider shear flow for which three kinds of driving are taken into 
account. The first is a shear stress
acting on the boundary, the other two are 
the cell-intrinsic active stresss and rearrangements.
Cell vertex model simulations have shown that directed cell rearrangements
may produced  self-driven shear flow \cite{Sato2015}.
In addition, the properties predicted by the following analysis will give 
opportunities to test the model in the future.

\paragraph{Assumptions} ~
We look for a solution with steady and uniform shear velocity gradient
in the form of
\begin{equation}
 \nabla \vec{v} = 
 \begin{pmatrix}
 0 & \dot{\gamma} \\
 0 & 0
\end{pmatrix} \,.
\end{equation}
In the incompressible case,  $\sigma'_{\mathrm{e}}$, $\sigma'_{\mathrm{p}}$, and $D_{\mathrm{r}}$ are given as follows
\begin{align}
\sigma'_{\mathrm{e}} &= \left(\frac{\gamma_0}{\sqrt{2\Tr{M}}} + 2\pi \kappa_0 \right) \frac{M'}{M_0} 
=
\Gamma(c) \sinh (c) \, \Theta \\
\sigma'_{\mathrm{p}} 
&= \eta D'  - \nu_1\sigma'_{\mathrm{e}} - \zeta_2 \Delta \mu  M'  - \zeta_{\mathrm{Q}} \Delta \mu Q \\
D_{\mathrm{r}} 
&= \nu_1 D'  + \eta^{-1}_1 \sigma'_{\mathrm{e}} -\beta_2 \Delta \mu M'- \beta_{\mathrm{Q}} \Delta \mu Q
\end{align}
Here, for simplicity, we omit possible dependences of the coefficients 
$\chi^{ss}$, $\chi^{sr}$, $\chi^{rs}$, and $\chi^{rr}$ on $M'$.
With an orientation along $\vec{n} = (\cos \phi, \sin \phi)^T$, 
the external signal reads
\begin{equation}
Q = \frac{1}{2}
\begin{pmatrix}
\cos 2 \phi & \sin 2 \phi \\
\sin 2 \phi & - \cos 2 \phi
\end{pmatrix} \,.
\end{equation}
Writing $D_{\mathrm{r}}$  as
\begin{equation}
 D_{\mathrm{r}} = 
 \begin{pmatrix}
 d_r & \delta_r \\
 \delta_r & -d_r
\end{pmatrix} \,,
\end{equation}
$d_r$ and $\delta_r$ are given as follows
\begin{align}
d_r &=  
\left( \frac{\Gamma(c)}{\eta_1}   - \beta_2\Delta \mu \, M_0 \right) 
\sinh c \, \cos 2\theta - \frac{\beta_{\mathrm{Q}} \Delta \mu}{2}   \, 
\cos 2 \phi  \,,\label{eq:dr}\\
\delta_r &= 
 \frac{\nu_1}{2} \dot{\gamma} + \left( \frac{\Gamma(c)}{\eta_1} 
-\beta_2\Delta \mu \, M_0 \right) \sinh c \, \sin 2\theta  -
\frac{\beta_{\mathrm{Q}} \Delta \mu }{2}  \, \sin 2 \phi 
 \,. \label{eq:dr2}
\end{align}

\paragraph{Kinematics}~
The kinematic equation at steady state ($\dot{M} = (\nabla \vec{v} -D_{\mathrm{r}})M+M(\nabla \vec{v} -D_{\mathrm{r}})^T=0$) leads to
\begin{align}
 \left( \cosh c +  \sinh c \cos 2\theta \right)d_r - \sinh c \sin2 \theta (\dot{\gamma} - \delta_r) = 0 \label{eq:k1}\\
 \cosh c (\dot{\gamma}-2\delta_r) - \dot{\gamma}\cos 2 \theta  \sinh(c) = 0 \label{eq:k2}\\
 \left( \cosh c -  \sinh c \cos 2\theta \right)d_r - \sinh c \sin 2\theta \delta_r = 0 \label{eq:k3}
\end{align}
One of these three equations is not independent of the others, because of the constraint that $|M|$ is constant.
With some calculation, we derive two independent equations
\begin{align}
 \dot{\gamma}-2\delta_r = \dot{\gamma}\cos 2\theta \tanh(c) \\
 2d_r  =  \dot{\gamma} \tanh(c ) \sin 2\theta
\end{align}
By substituting Eqs.~\eqref{eq:dr} and \eqref{eq:dr2}, we reach the equations
\begin{align}
2 \left( \frac{\Gamma(c)}{\eta_1}   - \beta_2\Delta \mu \, M_0 \right) 
\sinh c \, \sin 2\theta
 - \beta_{\mathrm{Q}} \Delta \mu \, \sin 2 \phi &=
(1-\nu_1-\tanh c \, \cos 2\theta ) \, \dot{\gamma} 
 \label{eq:EQ2}\\
2 \left( \frac{\Gamma(c)}{\eta_1}   - \beta_2\Delta \mu \, M_0 \right) 
\sinh c \, \cos 2\theta
 - \beta_{\mathrm{Q}} \Delta \mu \, \cos 2 \phi
&=  \tanh c \,\sin 2\theta \,\dot{\gamma} 
\,,\label{eq:EQ3}
 \end{align}
which determine cell shape $(c, \theta)$ for a given  shear rate $\dot{\gamma}$.

\paragraph{Stress} 
The total stress $\sigma = \sigma'_{\mathrm{e}} + \sigma'_{\mathrm{p}} - pI$ reads
\begin{equation}
\sigma = (1-\nu_1) \, \Gamma(c) \sinh c \, \Theta + 
 \eta D'  - \zeta_2 \Delta \mu \, M'  
 - \zeta_{\mathrm{Q}} \Delta \mu \, Q -p I \,.
\end{equation}
The stress boundary condition at $y = +L$ (top) is given as 
$\sigma_{xy} = \sigma_{\mathrm{b}}$, where $\sigma_{\mathrm{b}}$ is the force per 
unit length applied at the boundary to drive the shear flow. 
This condition reads
\begin{equation}
\sigma_{\mathrm{b}} + \frac{\zeta_{\mathrm{Q}} \Delta \mu}{2} \, \sin 2 \phi  = 
\frac{\eta}{2} \dot{\gamma} +
(1-\nu_1)\Gamma(c) \, \sinh c \, \sin 2\theta 
  -\zeta_2 \Delta \mu \, M_0 \, \sinh c \, \sin 2\theta 
\label{eq:ForceBalanceAtBounary}
\end{equation}
The active stress $\zeta_{\mathrm{Q}} \Delta \mu \, \sin 2 \phi/2 $ 
plays a role equivalent to  the external driving stress
$\sigma_{\mathrm{b}}$ 
in the sense that it shifts $\sigma_{\mathrm{b}}$ by a constant as 
$ \sigma'_{\mathrm{b}} = \sigma_{\mathrm{b}} - \zeta_{\mathrm{Q}} \Delta \mu \, \sin 2 \phi/ 2$.

\paragraph{Shear rate} 
From Eqs.~\eqref{eq:EQ2}, \eqref{eq:EQ3} and \eqref{eq:ForceBalanceAtBounary}, 
we can evaluate how the shear rate $\dot{\gamma}$ depends on the driving stress
$\sigma_{\mathrm{b}}$.
\begin{equation}
 \dot{\gamma} = 2  \frac{\sigma_{\mathrm{b}} + 
 \Delta \mu \,
\left( (\zeta_{\mathrm{Q}}  - (1-\nu_1)\eta_1 \beta_{\mathrm{Q}})  \sin 2 \phi 
+  ( \zeta_2- 2 \eta_1 (1-\nu_1) \beta_2) M_0\,\sinh c \, \sin 2 \theta  \right)
}{\eta + (1-\nu_1) \eta_1 [1-\nu_1-  \tanh c \cos 2\theta]}
\label{eq:gammadot}
\end{equation}
For $\sigma_{\mathrm{b}} = \zeta_{\mathrm{Q}} = \beta_2 = \zeta_2 = 0$,
\eqref{eq:gammadot} shows that oriented active rearrangements suffice 
to generate shear flow, as shown using the CVM in \cite{Sato2015}
with an orientation along $\vec{n} = (-1/\sqrt{2}, 1/\sqrt{2})$, 
with $\phi = 3\pi/4$, and an external signal 
\begin{equation}
Q = \frac{1}{2}
\begin{pmatrix}
0 & -1 \\
-1 & 0
\end{pmatrix} \,.
\end{equation}

\paragraph{Shear thinning}
For $\Delta \mu = 0$, $ \eta_{\mathrm{eff}} = 2 \sigma_{\mathrm{b}}/\dot{\gamma}$ 
is not constant, 
indicating that the tissue is a non-Newtonian material (Fig.~5(b-c) in the main text).
As $\sigma_{\mathrm{b}}$ and accordingly $\dot{\gamma}$ increase, $\eta_{\mathrm{eff}}$ converges to $\eta_{\mathrm{eff}}^{\infty} = \eta$
(Fig.~5(c)). 
This convergence occurs at the rate $\eta_{\mathrm{eff}} - \eta \sim \dot{\gamma}^{-2}$,
as shown in the numerical calculation (Fig.~\ref{fig:SheaThinning_Diff}).

To understand this dependence of $\eta_{\mathrm{eff}}$ on $\dot{\gamma}$, 
let us consider
Eqs.~\eqref{eq:EQ2} and \eqref{eq:EQ3} with $\Delta \mu = 0$.
\begin{align}
2 \Gamma(c) \sinh c \, \sin 2\theta &= \eta_1 (1-\nu_1-\tanh c \, \cos 2\theta ) \, \dot{\gamma}  \label{eq:EQ_A} \\
2 \Gamma(c) \sinh c \, \cos 2\theta &=  \eta_1 \tanh c \,\sin 2\theta \,\dot{\gamma}  \, \label{eq:EQ_B}
 \end{align}
For the right hand sides of these equations to remain finite 
in the limit $\dot{\gamma} \to \infty$,  $c$ and $\theta$ converge 
to $c \to c^{\infty}$ and $\theta \sim \dot{\gamma}^{-1} \to 0$,
respectively, where $c^{\infty}$ is a solution of the following equation:
\begin{equation}
  \label{eq:cinfty:1}
 \tanh c^{\infty} = 1-\nu_1 \,. 
\end{equation}
 Considering small deviations $c = c^{\infty} - \Delta c$ and 
$\theta = \Delta \theta$, Eq.~\eqref{eq:EQ_A} reads
\begin{equation}
  \label{eq:cinfty:2}
4\Gamma(c^{\infty}) \sinh c^{\infty}\Delta \theta = \frac{\Delta c}{\cosh^2 c^{\infty}} \dot{\gamma} \,,
\end{equation}
thus $\Delta c$ is of the order of $ \Delta c \sim \dot{\gamma}^{-2}$. 
The difference $\eta_{\mathrm{eff}} - \eta$
for large $\dot{\gamma}$ is evaluated from Eq.~\eqref{eq:gammadot}, as 
\begin{align}
 \eta_{\mathrm{eff}} - \eta & = (1-\nu_1) \eta_1 \left[ 1-\nu_1 - \tanh c \cos 2 \theta\right] \nonumber \\
 & \sim  (1-\nu_1) \eta_1 \frac{\Delta c}{\cosh^2 c^{\infty}}  \nonumber \\
 & \sim   \eta_1 \nu_1 (1-\nu_1) (2-\nu_1) \, \Delta c\,,
\end{align}
which is of the order of  $\dot{\gamma}^{-2}$.

%\break
%%%%% References %%%%%

\break

%%%%%%%%%%%%%%%%%

%%%%% Applications 2%%%%%%%
%%%%%%%%%%%%%
%%Fig. 4
%%%%%%%%%%%%
%%Fig.5

\break

\renewcommand{\figurename}{Fig. S\!\!}
\setcounter{figure}{0}

\begin{figure}[bt]
\centering
 \resizebox{0.5\textwidth}{!}{%
   \includegraphics{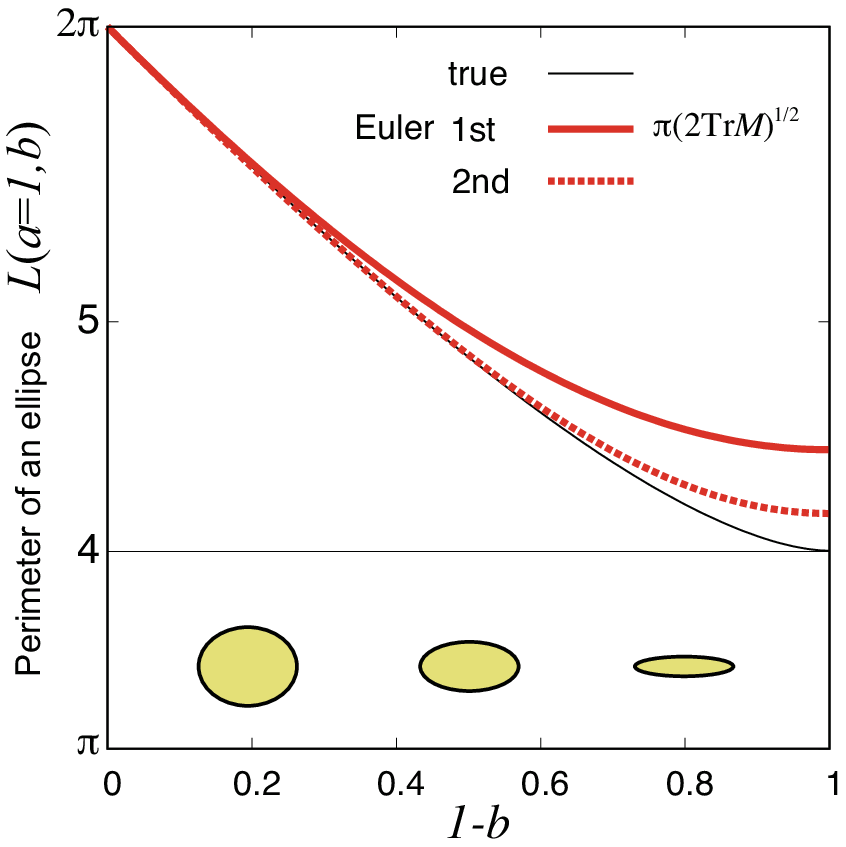} 
  }
  \caption{ \label{fig:EulerApprox}
	Euler approximation of the perimeter $L$ of an ellipse of semi-axes 
$a = 1$ and $b \in [0,1]$ as a function of $1-b$ at first and  second order
is compared to the exact value. The isotropic case corresponds to $a = b$,
or $1-b = 0$.
}
\end{figure}

\begin{figure}[h!]
\centering
 \resizebox{0.74\textwidth}{!}{%
   \includegraphics{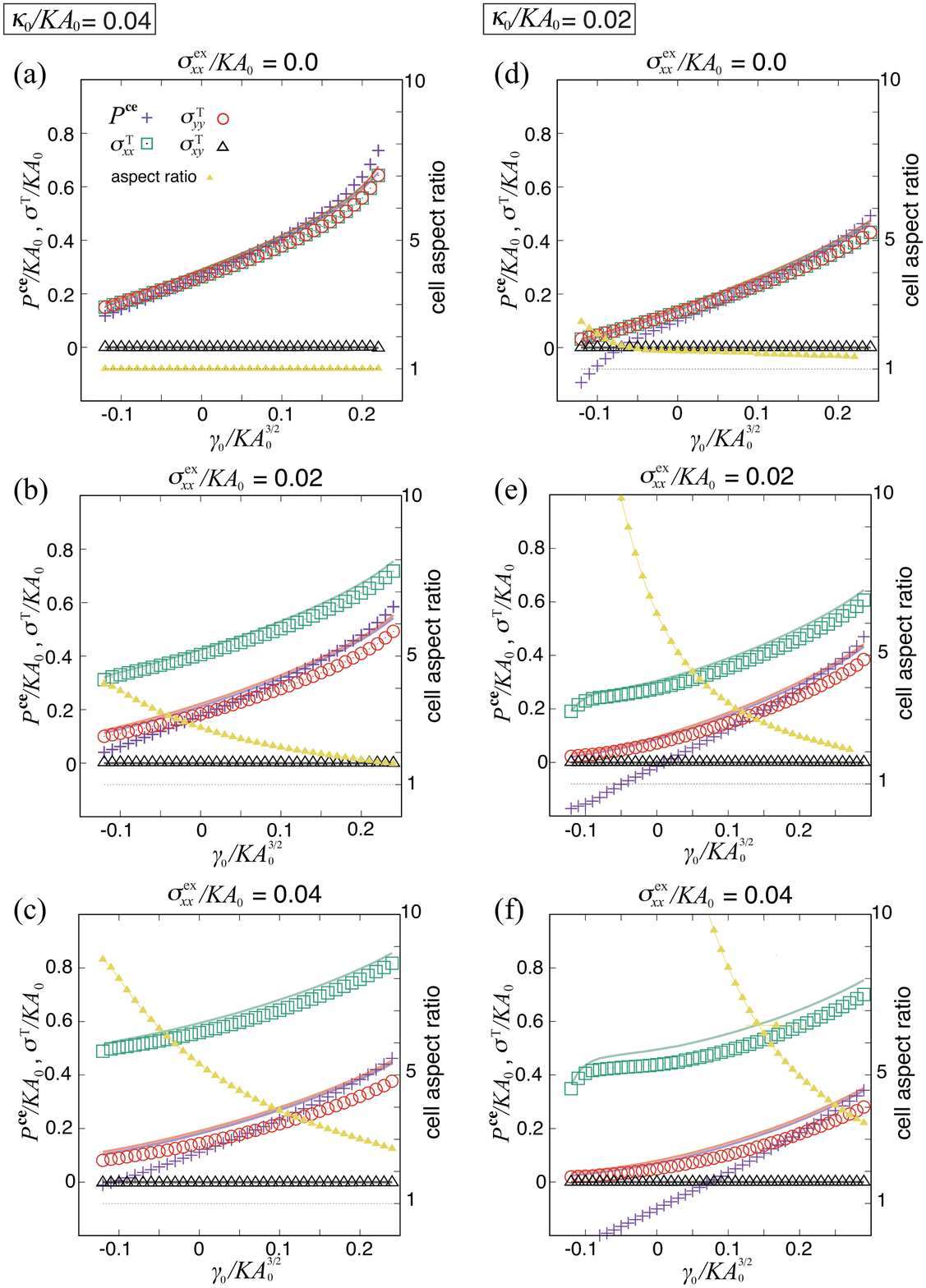} 
  }
  \caption{ \label{fig:CVM}
	Macroscopic stress expressions calculated from coarse-grained cell 
	shape tensor $M$ (symbols)  are compared with the true ones (solid lines) 
	obtained by CVM simulations (left vertical axis), as a function of
	the non-dimensionalized parameter  $\gamma_0/KA_0^{3/2}$,
	with $\kappa_0/KA_0 = 0.02$ (right column) and $\kappa_0/KA_0 = 0.04$ 
	(left column).  $P^{\mathrm{ce}}$ and $\sigma^{\mathrm{T}}$ are
	the pressure and stresses originating from cell elasticity and cell junction 
	tensions, respectively. Yellow triangles denote the mean cell shape 
	aspect ratio (right vertical axis), equal to $\exp(2c)$ in terms of the
	cell shape anisotropy.
	Components of the external stress are set as $\sigma^{\mathrm{ex}}_{xy} 
	=\sigma^{\mathrm{ex}}_{yy} = 0$, with (from left to right column) 
	$\sigma^{\mathrm{ex}}_{xx} = 0.0$, $0.02$, $0.04$.
  }
\end{figure}

\begin{figure}[t!]
\centering
 \resizebox{0.6\textwidth}{!}{%
 \includegraphics{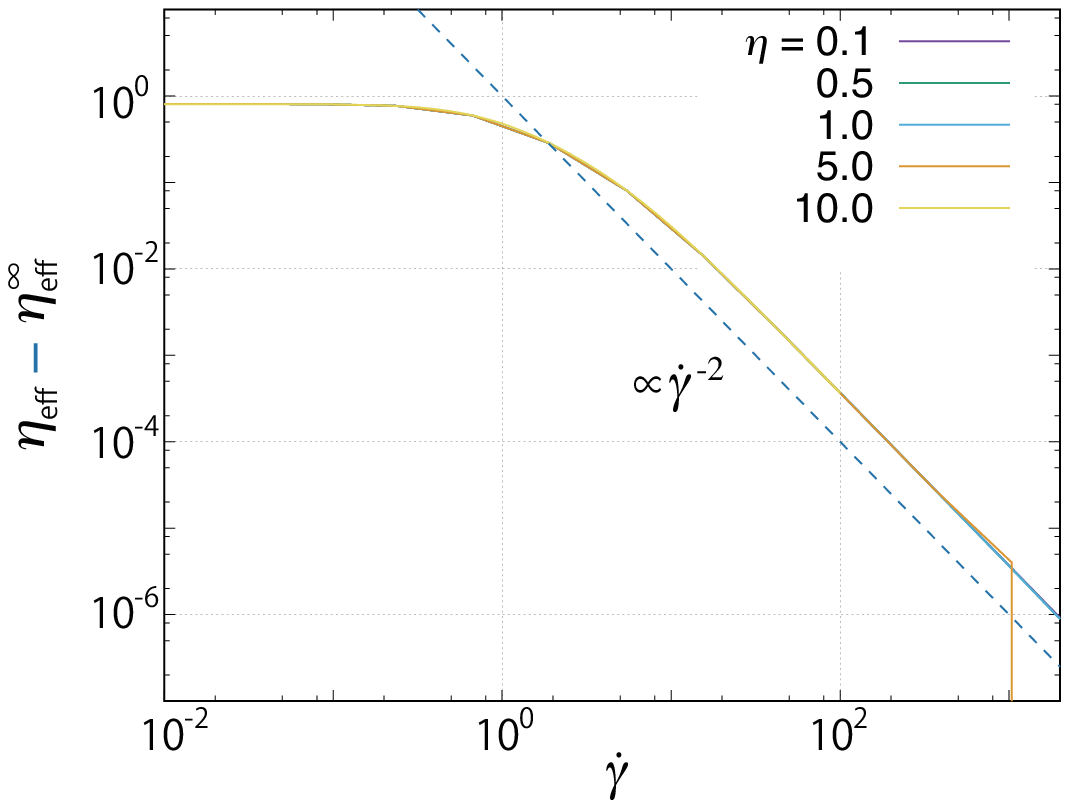}
  }
  \caption{ \label{fig:SheaThinning_Diff}
Log-log plot of the difference 
$\eta_{\mathrm{eff}} - \eta_{\mathrm{eff}}^{\infty}$ as a function of the shear rate 
$\dot\gamma$. The dashed line corresponds to
$\eta_{\mathrm{eff}} - \eta_{\mathrm{eff}}^{\infty} \propto \dot{\gamma}^{-2}$.
}
\end{figure}

\end{document}